\documentclass[aps,pre,twocolumn,groupedaddress]{revtex4-1}
\usepackage{fullpage}
\usepackage{mathtools}
\usepackage[caption=false]{subfig}
\usepackage{gensymb}
\usepackage{wrapfig}
\usepackage{color}
\usepackage{lineno}
\usepackage[bookmarks=false]{hyperref}
\usepackage[nameinlink]{cleveref}
\crefname{equation}{}{}
\crefname{figure}{Fig.}{Figs.}
\crefname{table}{Table}{Tables}

\begin{document}

\title{Numerical analysis of the lattice Boltzmann method for simulation of linear acoustic waves}

\author{Dattaraj B. Dhuri}
\email[]{dattaraj.dhuri@tifr.res.in}
\affiliation{Tata Institute of Fundamental Research (TIFR), Mumbai, India}
\author{Shravan M. Hanasoge}
\affiliation{Tata Institute of Fundamental Research (TIFR), Mumbai, India}
\author{Prasad Perlekar}
\affiliation{TIFR Centre For Interdisciplinary Sciences, Hyderabad, India}
\author{Johan O.A. Robertsson}
\affiliation{ETH Z\"urich, Institute of Geophysics, Z\"urich, Switzerland}

\date{\today}

\begin{abstract}
We analyse a linear lattice Boltzmann (LB) formulation for simulation of linear acoustic wave propagation in heterogeneous media. We employ the single-relaxation-time Bhatnagar-Gross-Krook (BGK) as well as the general multi-relaxation-time (MRT) collision operators. By calculating the dispersion relation for various 2D lattices, we show that the D2Q5 lattice is the most suitable model for the linear acoustic problem. We also implement a grid-refinement algorithm for the LB scheme to simulate waves propagating in a heterogeneous medium with velocity contrasts. Our results show that the LB scheme performance is comparable to the classical second-order finite-difference schemes. Given its efficiency for parallel computation, the LB method can be a cost effective tool for the simulation of linear acoustic waves in complex geometries and multiphase media.
\end{abstract}


\maketitle

\section{Introduction \label{sec:introduction}}
Over the past three decades, the lattice Boltzmann (LB) method has been established alongside conventional CFD methods as an efficient scheme for the numerical solution of partial differential equations. A variety of complex flow problems \cite{BENZI1992145,doi:10.1146/annurev-fluid-121108-145519} have been tackled successfully, even within the simplest Bhatnagar-Gross-Krook (BGK) framework \cite{PhysRev.94.511}. It has proven particularly useful for applications involving flows through porous and multi-phase media \cite{doi:10.1146/annurev.fluid.30.1.329,illner2007lattice,succi2001lattice,PhysRevLett.112.014502}. The method has been extended well beyond hydrodynamics to solve the governing equations of magnetohydrodynamics \cite{PhysRevLett.67.3776}, acoustic and electromagnetic wave propagation \cite{Chopard1999115,0295-5075-96-1-14002}, fracturing in solids \cite{9780511549755}, the Schr\"odinger equation \cite{PhysRevE.75.066704} etc. With a generalized multi-relaxation-time (MRT) collision operator \cite{el00313,dHumi2002,PhysRevE.61.6546}, the LB scheme can be fine tuned to more accurately model the physical problem at hand.

The LB method has several distinct advantages over conventional CFD approaches \cite{doi:10.1146/annurev.fluid.30.1.329}. The LB equation comprises a single first-order differential equation, which, in the asymptotic limit (Chapman-Enskog expansion procedure \cite{chapman1970,epic23739}), produces the macroscopic equations for mass and momentum conservation. Only immediate neighbour lattice sites interact in LB, so that the computation is highly suitable for a parallel-computing implementation. Imposing boundary conditions in LB is relatively simple, which makes it suitable to handle complex geometries.

In the current work, we discuss an application of LB to model linear wave propagation in seismology. For simplicity and clarity in understanding numerical performance, we consider a seismic wave equation of only acoustic waves (P-waves) with applications in e.g. helioseismology \cite{Gizon2010}. The propagation medium for these seismic waves can be highly heterogeneous, comprising a mixture of multi-phase components in a porous environment. The geometry of the physical domain of interest may be complicated. The distinct success of the LB method for multi-phase and porous flows combined with its efficiency in handling complex boundaries makes it useful to investigate how LB simulations of acoustic waves compare with standard numerical schemes --- such as finite-differences --- employed in exploration seismology.

LB schemes have been analysed in the context of simulation of acoustic waves in prior literature. In \cite{marie2009comparison}, Mari\'e et al. investigate sources of errors in LB dispersion as compared to Navier-Stokes. In \cite{xu2011optimal}, Xu et al. propose an optimization strategy to minimize dispersion errors in the MRT-LB scheme. In both these studies, a conventional LB scheme with a non-linear equilibrium distribution, which yields the Navier-Stokes equation, i.e., on performing the Chapman-Enskog expansion, is used \cite{epic23739}. Hence, the resulting wave equation retains some non-linearity as well as viscous dissipation and does not faithfully represent the linear inviscid wave equation \cref{eq:linearpresswave}. Instead, a straightforward way to simulate the inviscid linear wave equation is to use the linear LB scheme proposed by Chopard \cite{Chopard1999115}. The linear LB scheme utilises the linear equilibrium distribution function \cref{eq:linear eqm dist} and appropriately recovers only the linear inviscid part of the wave equation (see Appendix \ref{sec:Appendix B}). In this work, we analyse the linear LB scheme for the simulation of acoustic waves. Quantifying the numerical ability of a scheme is best done on a fully linear equation, e.g. \cite{lele1992compact}. Indeed, our work is relevant in this regard since it highlights unexpected strengths and weaknesses of the LB method.

Few authors have considered the linear LB scheme for the simulation of waves. Viggen \cite{ViggenPhd05,viggen2013sound,PhysRevE.87.023306} provide a detailed account of the application of LB scheme for acoustic waves.  In particular, \cite{ViggenPhd05} presents a derivation of the dispersion relation for the discrete-velocity Boltzmann equation using the linear equilibrium distribution function and compares it with the Navier-Stokes equation. As we will show later, the linear equilibrium distribution function can be written as a linear combination of single-particle distribution functions. Thus we can directly obtain the numerical dispersion relations without additional linearisation procedures. We derive dispersion relations for the MRT-LB scheme on 2D lattices --- D2Q5 and D2Q9. We also study the numerical anisotropy of the dispersion relation on these 2D lattices. The dispersion relations for the BGK-LB scheme are obtained as a special case when all the relaxation parameters are identical.  (similar to finite-difference schemes) The numerical dispersion relation, as well as the stability of the LB scheme, are sensitive to the Courant number \cite{robertsson2011numerical}. A heterogeneous medium may display large variations in sound speed, and hence the local Courant number.  Thus we study LB dispersion relations at different Courant numbers and compare it with second- and fourth-order finite-difference schemes. Significant changes in sound speed result in corresponding variations in wavelength. A grid-refinement algorithm (based on \cite{PhysRevE.67.066707}) is presented to simulate waves with uniform grid resolution, i.e., the number of grid points that resolve the wave. Although our analysis is performed on a strictly linear problem, the numerical limits of LB that we have identified are also relevant for standard LB schemes.

In Section \ref{sec:seismic waves}, we briefly discuss the LB methodology and its application to seismic wave propagation. We describe the mathematical model for seismic waves and linear LB formulation. We also discuss the parameters which are important for performance of the LB scheme for the simulation of waves. In Section \ref{sec:LB dispersion analysis}, we derive the dispersion relation for the LB scheme and present LB dispersion curves on 2D lattices. We compare LB dispersion curves with exact as well as finite-difference schemes at different Courant numbers. In Section \ref{sec:LB grid refinement}, we give a detailed account of the grid-refinement algorithm used to model waves in heterogeneous media. Finally, in Section \ref{sec:numerical experiments}, we present results from simulations in homogeneous and heterogeneous media.

\section{Seismic waves \label{sec:seismic waves}} 
\subsection{Macroscopic equations}
Seismic P-waves in acoustic media can be described by a linear pressure wave equation with a source term
\begin{equation}\label{eq:linearpresswave}
\frac{1}{c_{s}^{2}(\textbf{x})}\partial_{t}^{2}p(\textbf{x},t)-\nabla^{2}p(\textbf{x},t)=\partial_{t}S(\textbf{x},t).
\end{equation}
This second-order equation derives from two coupled first-order equations for continuity 
\begin{equation}\label{eq:energy equation}
\partial_{t} p(\textbf{x},t)+c_{s}^{2}(\textbf{x}) \nabla \cdot \big[\rho_{0}(\textbf{x}) \textbf{v}(\textbf{x},t)\big]=c_{s}^{2}(\textbf{x})S(\textbf{x},t),
\end{equation}
and conservation of momentum
\begin{equation}\label{eq:momentum equation}
\partial_{t} \big[\rho_{0}(\textbf{x})\textbf{v}(\textbf{x},t)\big]+\nabla p(\textbf{x},t)=0.
\end{equation}
Here $p(\textbf{x}, t)$ is pressure fluctuation, $\textbf{v}(\textbf{x}, t)$ is velocity fluctuation, $S(\textbf{x}, t)$ is the scalar source of pressure fluctuations, and $c_{s}(\textbf{x})$ and $\rho_{0}(\textbf{x})$ are the prescribed temporally stationary sound speed and density of the background medium. The functional form of $c_{s}(\textbf{x})$ and $\rho_{0}(\textbf{x})$ are dictated by the heterogeneity of the medium and may thus be complicated. The pressure and the density fluctuations are related by a linearised ideal gas equation of state
\begin{equation}\label{eq:state}
p\left(\textbf{x},t\right)=c_s^2\left(\textbf{x}\right)\rho\left(\textbf{x},t\right).
\end{equation}
The sound speed $c_{s}(\textbf{x})$ can be modelled using an isentropic bulk modulus for the medium \cite{kruger2017}.

\subsection{The LB model \label{sec:lbmodel}} 
Kinetic theory describes the dynamics of a gas in terms of the distribution function, which is the probability density of finding a gas particle in a differential phase-space volume. The distribution function evolves as particles move (or in LB terminology, stream) and collide with each other. In the BGK-Boltzmann transport equation, collision is modelled through a linear operator with a single relaxation time
\begin{equation}\label{eq:Boltzmann transport}
\begin{aligned}[b]
\partial_{t}g(\textbf{x},\textbf{c},t)&+\textbf{c}\cdot\nabla g(\textbf{x},\textbf{c},t) \\ ={}& -(1/\tau)\big[g(\textbf{x},\textbf{c},t)-g^{eq}(\textbf{x},\textbf{c},t)\big],
\end{aligned}
\end{equation}
where $g(\textbf{x},\textbf{c},t)$ is the single-particle distribution function \cite{huang1987statistical}, $\tau$ is the relaxation time for particle collisions and $g^{eq}(\textbf{x},\textbf{c},t)$ is the equilibrium Maxwell-Boltzmann distribution.

A second-order discretisation of the BGK-Boltzmann transport equation \cref{eq:Boltzmann transport} gives the BGK-LB equation \cite{reider1995accuracy,PhysRevE.56.6811}
\begin{equation}\label{eq:BGK-LB}
\begin{aligned}[b]
g_{i}(\textbf{x}+\textbf{c}_{i}\delta t,t&+\delta	t)-g_{i}(\textbf{x},t)\\={}&-(1/\tau)\big[g_{i}(\textbf{x},t)-g_{i}^{eq}(\textbf{x},t)\big].
\end{aligned}
\end{equation}
Both position space and velocity space are discretised; i.e., only a finite number of microscopic velocities are allowed. This is achieved by mapping the computational domain onto a lattice. In  \cref{eq:BGK-LB}, $\textbf{c}_i$ is the $i^{th}$ microscopic velocity on the lattice and $g_{i}(\textbf{x},t)$ is the corresponding single-particle distribution function. For the LB scheme, the equilibrium Maxwell-Boltzmann distribution is truncated at the second order in velocity \cite{ref1}
\begin{equation}\label{eq:LB M-B}
\begin{aligned}[b]
 g_{i}^{eq}(\textbf{x},t)={}&\frac{w_{i}}{ c_s^2(\textbf{x})}\Big\{p(\textbf{x},t)+\rho_0(\textbf{x})[\textbf{c}_i\cdot\textbf{v}(\textbf{x},t)]\\&+\frac{\rho_0(\textbf{x})}{2c_s^2(\textbf{x})}[\textbf{c}_i\cdot\textbf{v}(\textbf{x},t)]^2\\&-\frac{\rho_0(\textbf{x})}{2}\vert\vert\textbf{v}(\textbf{x},t)\vert\vert^2\Big\},
\end{aligned}
\end{equation}
where $w_{i}$ is the lattice weight for the $i^{th}$ microscopic velocity and $c_s(\textbf{x})$ is the lattice sound speed (see Section \ref{sec:LB parameters}). Here, the density and momentum fluctuations are obtained by taking zeroth and first microscopic velocity moments of the distribution function over the lattice velocity space,
\begin{equation}\label{eq:macroscopic density}
\rho (\textbf{x},t) = \sum_{i} g_{i}(\textbf{x},t) = \sum_{i} g_{i}^{eq}(\textbf{x},t) ,
\end{equation}
\begin{equation}\label{eq:macroscopic velocity}
\rho_0(\textbf{x})\textbf{v}(\textbf{x},t) =\sum_{i}\textbf{c}_{i} g_{i}(\textbf{x},t)= \sum_{i}\textbf{c}_{i} g_{i}^{eq}(\textbf{x},t).
\end{equation}
The pressure fluctuations $p(\textbf{x},t)$ is then obtained by using the equation of state \cref{eq:state}.
\begin{figure}
  \subfloat[]{\includegraphics[width= 0.5\columnwidth]{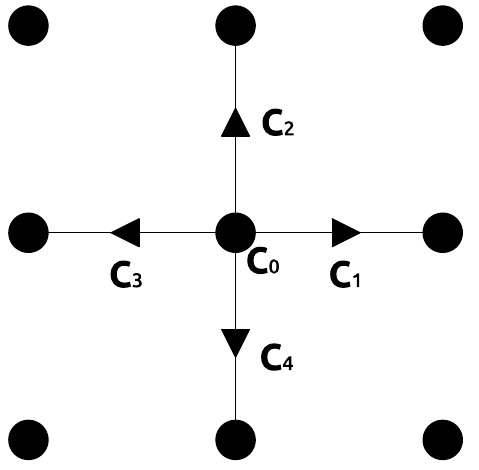}}\\
  \subfloat[]{\includegraphics[width= 0.5\columnwidth]{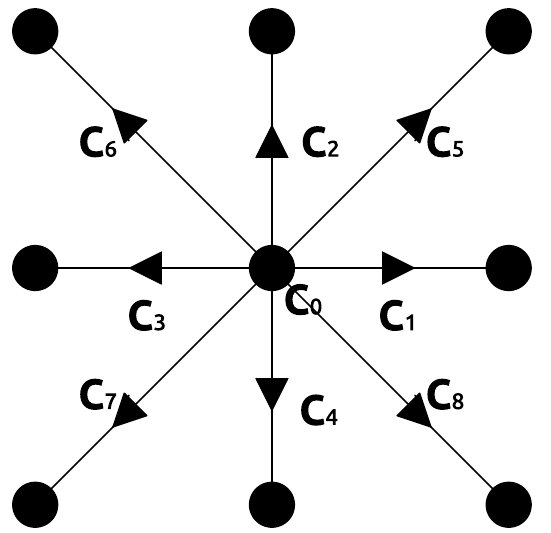}}
  \caption{\label{fig:lattices}Lattices used in simulation (a) D2Q5 lattice  (b) D2Q9 lattice}
\end{figure}
\begin{table}[b]
\centering
\begin{tabular}{|c|c|}
  \hline
  Lattice & Lattice velocities \\ \hline
  D2Q5 & \parbox[t]{5cm}{$c_0 = (0,0),$\\ $c_{1,3},c_{2,4} = (\pm 1,0),(0,\pm 1).$} \\ \hline
  D2Q9 & \parbox[t]{5cm}{$c_0 = (0,0),$\\ $c_{1,3},c_{2,4} = (\pm 1,0),(0,\pm 1),$\\$c_{5,7,6,8} = (\pm 1,\pm 1).$}\\ \hline
\end{tabular}
\caption{Lattice velocities}\label{tab:Lattice velocities}
\end{table}

Depending on the problem, various 2D and 3D lattices are used in LB. The general nomenclature for lattices is DnQm where n is the dimensions of position space and m is the number of microscopic velocities available at each lattice site. For simulating 2D acoustic waves, we have used two 2D lattices - D2Q5 and D2Q9, with 5 and 9 microscopic velocities respectively. These lattices are shown in \cref{fig:lattices} and the corresponding lattice velocities are given in \Cref{tab:Lattice velocities}. In both lattices, the microscopic velocity $c_0$ corresponds to the particle at rest.

In the MRT-LB scheme, the different components of the distribution function $\textbf{g}$ may relax to the equilibrium distribution function $\textbf{g}^{eq}$ with different relaxation parameters \cite{el00313}. The general LB equation is written for the distribution function column vector $\textbf{g}=\{g_{i}\}$ as,
\begin{equation} \label{eq:MRT-LB basic}
\begin{aligned}[b]
\textbf{g}(\textbf{x}+\textbf{c}\delta t,&t+\delta t)- \textbf{g}(\textbf{x},t)\\={}&- \rm{M}^{-1}S\ \big[\textbf{m}(\textbf{x},t)-\textbf{m}^{eq}(\textbf{x},t)\big].
\end{aligned}
\end{equation}
Here, the column vector $\textbf{m}$ consists of conserved and non-conserved velocity moments of the distribution function. The orthogonal matrix $\rm{M}$ transforms the distribution function vector into the moment vector as $\textbf{m}=\rm{M}\textbf{g}$. The conserved moments density and momentum fluctuations are given by equations \cref{eq:macroscopic density} and \cref{eq:macroscopic velocity} respectively. The non-conserved moments are at higher order in velocity and are calculated using the microscopic velocity set on a given lattice \cite{el00313,dHumi2002,PhysRevE.61.6546}. For the D2Q5 lattice, the moment column vector is
\begin{equation} \label{eq:D2Q5 moments}
\textbf{m}=\{\rho,\rho_0 v_x, \rho_0 v_y, e, p_{xx}\},
\end{equation}
and for the D2Q9 lattice,
\begin{equation} \label{eq:D2Q9 moments}
\textbf{m}=\{\rho,\rho_0 v_x, \rho_0 v_y, e, p_{xx}, \epsilon , q_x, q_y, p_{xy}\}.
\end{equation}
Second-order velocity moments $e$, $p_{xx}$ and $p_{xy}$ correspond to the energy, diagonal and off-diagonal components of stress tensor respectively and $\epsilon$, $q_x$ and $q_y$ are higher-order velocity moments on lattice \cite{PhysRevE.61.6546}.

The equilibrium distribution function vector  $\textbf{g}^{eq}$ transforms into the equilibrium moment vector $\textbf{m}^{eq}$ in moment space. Transformation matrices $\textup{M}$ for D2Q5 and D2Q9 lattices are specified in Appendix \ref{sec:Appendix A}. $\rm{S}$ is the diagonal relaxation matrix consisting of the inverse relaxation-times for the conserved as well non-conserved moments. For the D2Q5 lattice, $\textup{S}=\textup{diag}(s_{\rho},s_v,s_v,s_e,s_p)$ and for the D2Q9 lattice, $\textup{S}=\textup{diag}(s_{\rho},s_v,s_v,s_e,s_p,s_{\epsilon},s_q,s_q,s_p)$ (where diag refers to diagonal matrix). The various inverse relaxation-times --- $s$ --- are related to the macroscopic properties of the fluid which govern the hydrodynamics and kinetics (Section \ref{sec:LB parameters}). The values of these parameters can be fine tuned to suit the macroscopic dynamics of the physical system and also to improve the numerical stability of the model \cite{PhysRevE.61.6546}. The generalised LB equation \cref{eq:MRT-LB basic} can be expanded to obtain the MRT-LB equation in terms of the distribution column vector alone,
\begin{equation}
\begin{aligned}[b]\label{eq:MRT-LB}
\textbf{g}(\textbf{x}+\textbf{c} & \delta t,t+\delta t)- \textbf{g}(\textbf{x},t) \\ ={}&-\rm{M}^{-1}\rm{S}\rm{M}~\big[\textbf{g}(\textbf{x},t)-\textbf{g}^{eq}(\textbf{x},t)\big]\\ ={}&-\rm{C}~\big[\textbf{g}(\textbf{x},t)-\textbf{g}^{eq}(\textbf{x},t)\big].
\end{aligned}
\end{equation}
Here, $\rm{C}=\rm{M}^{-1}{S}{M}$ is the collision matrix. When all the inverse relaxation-times are identical, we recover the BGK-LB equation \cref{eq:BGK-LB}.

The linear wave equation \cref{eq:linearpresswave} can be modelled by introducing a source term in the MRT-LB equation \cref{eq:MRT-LB} \cite{PhysRevE.87.023306}. Writing component wise, the corresponding equation is
\begin{equation}\label{eq:MRT-LB linear acoustic}
\begin{aligned}[b]
g_{i}(\textbf{x}+\textbf{c}_{i}\delta t,&t+\delta	t)-g_{i}(\textbf{x},t)\\={}&-\sum_{j} \textit{C}_{ij}~\big[g_{j}(\textbf{x},t)-g_{j}^{eq}(\textbf{x},t)\big]\\&+ w_{i} \ S(\textbf{x},t),
\end{aligned}
\end{equation}
where $w_i$ are the lattice weights. Since the first-order macroscopic equations \cref{eq:energy equation,eq:momentum equation} are linear in fluctuating quantities $p(\textbf{x},t)$ and $\textbf{v}(\textbf{x},t)$, we truncate the equilibrium distribution at the linear term in $\textbf{v}(\textbf{x},t)$. Thus the numerical model of linear acoustic waves involves the following linear equilibrium-distribution function
\begin{equation}\label{eq:linear eqm dist}
g_{i}^{eq}(\textbf{x},t)=\frac{w_{i}}{c_{s}^{2}(\textbf{x})}\Big[p(\textbf{x},t)+  \rho_{0}(\textbf{x})v_{\alpha}(\textbf{x},t) c_{i\alpha}\Big].
\end{equation}
The MRT-LB equation \cref{eq:MRT-LB linear acoustic} along  with the linear equilibrium distribution \cref{eq:linear eqm dist} are collectively referred to as the linear LB scheme. The solution of \cref{eq:MRT-LB linear acoustic} using the equilibrium distribution function \cref{eq:linear eqm dist}, with proper choice of relaxation times, yields the numerical solution of the linear acoustic wave equation for pressure \cref{eq:linearpresswave} as well as velocity. We discuss relaxation times and other LB parameters in the next section.
 
\subsection{The LB parameters \label{sec:LB parameters}} 
The LB equation \cref{eq:MRT-LB linear acoustic} represents the governing equation for linear acoustic waves \cref{eq:linearpresswave} provided we correctly adjust the relaxation parameters --- $s$ for the MRT scheme and $\tau$ for the BGK scheme. The systematic procedure to obtain macroscopic conservation equations \cref{eq:energy equation,eq:momentum equation} from the LB equation is the Chapman-Enskog analysis \cite{chapman1970,epic23739}. For the BGK-LB scheme, the Chapman-Enskog expansion gives the kinematic viscosity on the lattice as \cite{epic23739}
\begin{equation}\label{sec:LB viscosity} 
\nu(\textbf{x})=c_{s}^{2}(\textbf{x})\left(\tau-\frac{1}{2}\right)\delta t.
\end{equation} 
The wave equation \cref{eq:linearpresswave} has no dissipation term i.e. kinematic viscosity is zero. This is achieved by setting the collision relaxation time $\tau=1/2$. This particular choice of $\tau$ thus recovers macroscopic equations \cref{eq:energy equation,eq:momentum equation} describing the linear acoustic wave \cite{9780511549755}.

For the MRT-LB scheme, the Chapman-Enskog analysis is carried out in moments space for each of the conserved and non-conserved moments. For the D2Q9 lattice, the analysis shows that the kinematic and bulk viscosities on the lattice may be set to zero with the choice $s_p=2$ and $s_e=2$ respectively \cite{PhysRevE.61.6546}.
For the D2Q5 lattice, setting $s_p=2$ and either $s_e=2$ or the lattice sound speed $c_s^2=1/2$, recovers linear macroscopic equations \cref{eq:energy equation,eq:momentum equation} (see Appendix \ref{sec:Appendix B}). Other relaxation parameters (i.e. apart from $s_p$ , $s_e$) do not affect the hydrodynamics of the problem. However, in the case of the D2Q9 lattice, any choice except $s_e=2$ and $s_\epsilon =2$ causes numerical instability. The origin of these instabilities may be revealed by an analysis for the linear MRT-LB scheme similar to one performed in \cite{Dellar2003351}.

In order to obtain the full Navier-Stokes equation from the LB equation, it is essential that microscopic velocity moments of lattice weights up to the fourth order are identical to that of the Maxwell-Boltzmann distribution with zero-mean velocity \cite{epic23739,frisch1986lattice}. However, the Chapman-Enskog expansion shows that, for the linear LB scheme that we use, velocity moments of lattice weights up to the second order are relevant for obtaining the macroscopic conservation Eqs. \cref{eq:energy equation,eq:momentum equation} \cite{Chopard1999115}. The constraints on lattice weights for the linear LB scheme are (all odd-order moments vanish)
\begin{equation}\label{eq:LB lattice moments}
\begin{split}
\sum_{i} w_{i}(\textbf{x}) &=1,\\
\sum_{i} w_{i}(\textbf{x})c_{i\alpha}c_{i\beta} &=c_{s}^{2}(\textbf{x})\delta_{\alpha \beta}.\\
\end{split}
\end{equation}
In the Chapman-Enskog expansion, the second constraint above determines proportionality between density and the pressure term in the momentum equation (see equations \cref{eq:appceo1c12} and \cref{eq:appceo2c12}) and hence the local sound speed $c_s(\textbf{x})$. The Chapman-Enskog expansion also shows that the local lattice sound speed may be controlled by spatially adjusting the rest-particle lattice weight $w_0$ \cite{Chopard1999115}. This is achieved by setting
\begin{equation}\label{eq:0th lattice weight}
w_{0}(\textbf{x})=1-\eta(\textbf{x})^{2},
\end{equation}
where $\eta(\textbf{x})=c_{s}(\textbf{x})/c_{s\max} \leq 1$ , where $c_{s\max}$ is the maximum sound speed in the medium. In addition, because of lattice symmetry, not all lattice weights are different. For instance, in the D2Q5 lattice, we must have $w_1=w_2=w_3=w_4$. Also, lattice weights cannot be negative. Thus, given the value of $\eta(\textbf{x})$, Eqs. \cref{eq:LB lattice moments,eq:0th lattice weight} determine the lattice parameters --- weights $w_i(\textbf{x})$ and sound speed $c_s(\textbf{x})$.

Using \cref{eq:linear eqm dist}, lattice weight constraints \cref{eq:LB lattice moments} and transformation matrix M (see Appendix \ref{sec:Appendix A}) for LB lattices, second- and higher-order velocity moments of the linear equilibrium distribution can be calculated. Thus, for the D2Q5 lattice, we get
\begin{equation}\label{eq:higher moments e}
\begin{aligned}[b]
e^{(0)} ={}& -4g_{0}^{(0)} + g_{1}^{(0)} + g_{2}^{(0)} + g_{3}^{(0)} + g_{4}^{(0)} \\={}& -4\rho + 10\rho c_s^2,
\end{aligned}
\end{equation} 
and
\begin{equation}\label{eq:higher moments pxx}
p_{xx}^{(0)} =  g_{1}^{(0)} - g_{2}^{(0)} + g_{3}^{(0)} - g_{4}^{(0)} = 0.
\end{equation} 
Similarly, higher-order velocity moments of the linear equilibrium distribution on the D2Q9 lattice are calculated.

\section{Dispersion analysis \label{sec:LB dispersion analysis}} 
\subsection{Grid Resolution (R) \label{sec:LB grid res}} 
In a numerical simulation, $\delta t \sim \delta x$ (acoustic scaling) and $\delta x \propto N^{-1/3}$ in 3D where N is the total number of grid-points. Hence, total number of time steps scale with $N^{1/3}$. The cost of a time step scales linearly with the total number of grid-points $N$. Hence, the cost of the numerical simulation scales as $\mathcal{O}\left(N^{4/3}\right)$ in 3D. Therefore, an important performance criterion of a numerical technique used to model wave propagation is the accuracy of the solution at low grid resolution. A natural way to define the grid resolution is to consider the number of grid points required to resolve the characteristic wavelength in the problem. Thus, the grid resolution is given by $R=(\lambda_c/\delta x)$, where $\lambda_c$ is the characteristic wavelength. The characteristic wavelength can, for instance, be the shortest wavelength in the problem. 

\begin{figure}[ht] 
\centering
\includegraphics[width=0.35\columnwidth]{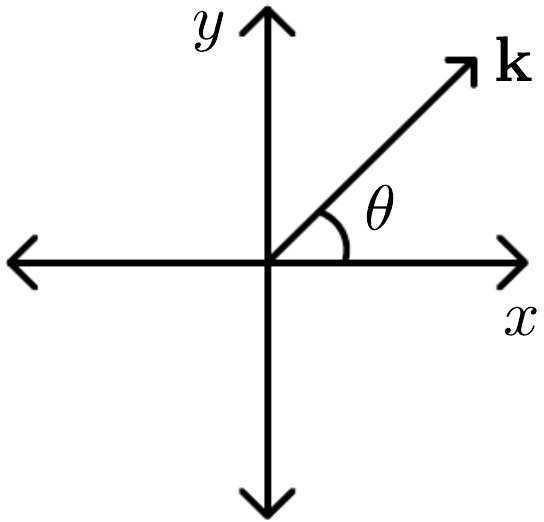} 
\caption{\label{fig:propagation direction} Direction of propagation for plane waves on the lattice. The numerical dispersion relation strongly depends on the angle $\theta$ along which the waves are propagating.}
\end{figure}
\subsection{LB dispersion relation \label{sec:LB dispersion relation}} 

The dispersion relation captures the connection between the vector wave-number $\textbf{k}$ and the frequency $\omega$ of the wave as captured by the numerical scheme. For waves travelling in homogeneous, dissipation-free media, $\omega$ and $k$ are proportional to each other in all directions i.e. there is no dispersion.  However, a numerical scheme used to simulate these waves may introduce artificial dispersion and attenuation. We analyse the dispersion characteristics of the MRT-LB scheme for D2Q5 and D2Q9 lattices by studying the response of the scheme to plane waves in homogeneous media i.e $c_s(\textbf{x}),\rho_0(\textbf{x}) = \textrm{constant}$. 

We can express the equilibrium distributions $\textbf{g}^{eq}$ as a linear combination of the distribution functions $\textbf{g}$ using \cref{eq:macroscopic density,eq:macroscopic velocity,eq:linear eqm dist}. This relation for the D2Q5 lattice is given by
\begin{equation}\label{eq:D2Q5 dist to eqm dist}
g_i^{eq}=w_i\Big\{\sum_{j} g_j+\big(1/c_s^2\big)\big[c_{ix}(g_1-g_3)+c_{iy}(g_2-g_4)\big]\Big\},
\end{equation}
and for the D2Q9 lattice,
\begin{equation}\label{eq:D2Q9 dist to eqm dist}
\begin{aligned}[b]
g_i^{eq}={}&w_i\Big\{\sum_{j} g_j+\big(1/c_s^2\big)\big[c_{ix}(g_1-g_3+g_5-g_6-g_7+g_8)\\&+c_{iy}(g_2-g_4+g_5+g_6-g_7-g_8)\big]\Big\}.
\end{aligned}
\end{equation}
Using these relations, equation \cref{eq:MRT-LB} can be rewritten as
\begin{equation}\label{eq:MRT-LB vector}
\begin{aligned}[b]
\textbf{g}(\textbf{x},t+\delta t)&- \textbf{g}(\textbf{x}-\textbf{c} \delta t,t) \\={}&- \textrm{M}^{-1}\textrm{S}\textrm{M}~\textrm{A}\textbf{g}(\textbf{x}-\textbf{c} \delta t,t),
\end{aligned}
\end{equation}
where for the D2Q5 lattice
\begin{widetext}
\[
\rm{A}=
  \begin{bmatrix}
    1-w_0 & -w_0 & -w_0 & -w_0 & -w_0 \\
    -w_1 & 1-w_1(1+1/c_s^2) & -w_1 & -w_1(1-1/c_s^2) & -w_1 \\
    -w_2 & -w_2 & 1-w_2(1+1/c_s^2) & -w_2 & -w_2(1-1/c_s^2) \\
    -w_3 & -w_3(1-1/c_s^2) & -w_3 & 1-w_3(1+1/c_s^2) & -w_3 \\
    -w_4 & -w_4 & -w_4(1-1/c_s^2) & -w_4 & 1-w_4(1+1/c_s^2)
  \end{bmatrix}.
\]
\end{widetext}
Similarly, matrix $\textup{A}$ for the D2Q9 lattice can be constructed.
\begin{figure*} 
\centering
\includegraphics[width=0.8\linewidth]{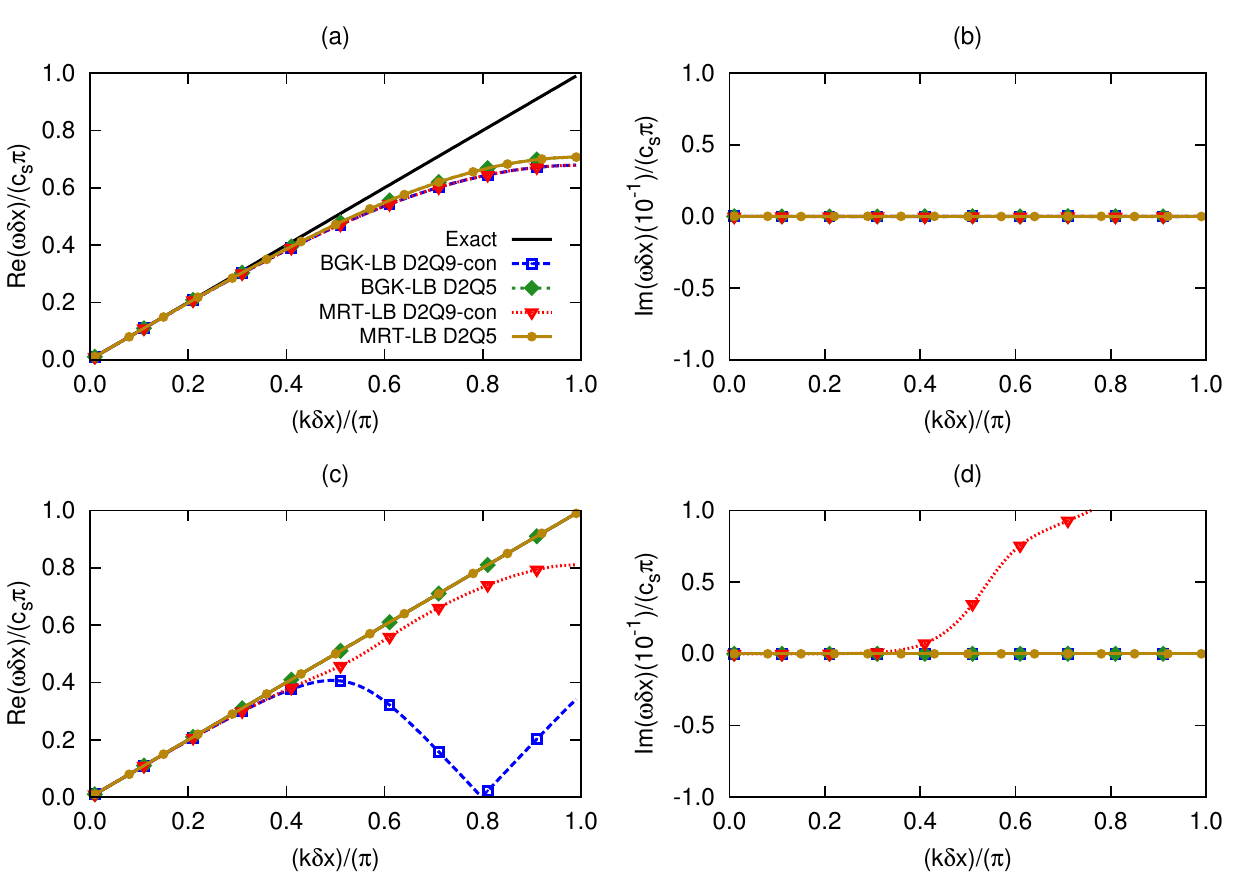} 
\caption{The BGK- and MRT-LB numerical dispersion and attenuation curves for D2Q5 and D2Q9 lattices. The real/imaginary component of the non-dimensional frequency $(\omega \delta x)/c_s\pi$, is plotted against the non-dimensional wave number $(k \delta x)/\pi$. Panels (a) and (c) show numerical dispersion for propagation along $\theta=0\degree$ and $\theta=45\degree$ respectively and panels (b) and (d) show numerical attenuation for propagation along $\theta=0\degree$ and $\theta=45\degree$ respectively. Lattice weights used for D2Q5 are $w_0=0, w_1=1/4$. For D2Q9, conventional lattice weights $w_0=4/9, w_1=1/9, w_5=1/36 $ are used (curves marked D2Q9-con). For the MRT implementation, relaxation parameters $s_p$ and $s_e$ are set to $2$ for both lattices.  For D2Q9, relaxation parameter $s_\epsilon$ is also set to $2$ and $s_q$ is set to $1$. Relaxation parameters for the conserved moments $s_\rho$ and $s_v$ can be set to any value as it does not affect the hydrodynamics. The D2Q5 dispersion curves for the BGK- and MRT-LB schemes are identical (better than D2Q9). The legend in the first plot also applies to the rest of the plots.}
\label{fig:MRT-LB dispersion} 
\end{figure*} 
The distribution function is initialized such that macroscopic variables $p(\textbf{x},t)$ and $\textbf{v}(\textbf{x},t)$ evolve in time as a plane wave. This is achieved by initializing the $j^{th}$ component of the distribution function (in Fourier space) as
\begin{equation}\label{eq:wave distribution}
g_j(\textbf{k},\omega) = w_j\,( 1 - \textbf{c}_j\cdot\textbf{k}/\omega)\,\exp[\iota(\textbf{k}\cdot\textbf{x} - \omega t)].
\end{equation}
Note that the coefficient of these distribution function plane waves is chosen such that we recover initial pressure and velocity plane waves consistent with equations \cref{eq:energy equation} and \cref{eq:momentum equation} from equations \cref{eq:macroscopic density} and \cref{eq:macroscopic velocity}.  Substituting \cref{eq:wave distribution} in the Fourier transform of \cref{eq:MRT-LB vector} gives us the dispersion relation for the MRT-LB scheme with a linear equilibrium distribution;
\begin{equation}\label{eq:MRT-LB dispersion}
\begin{aligned}[b]
\exp&(-\iota\omega \delta t) \textbf{g}(\textbf{k},\omega) \\={}&\Big(1- \textup{M}^{-1}\textup{S}\textup{M}\textup{A}\Big)\textup{diag}[\exp(-\iota\textbf{k}\cdot \textbf{c}_j \delta t)] \textbf{g}(\textbf{k},\omega),
\end{aligned}
\end{equation}
where $\textbf{g}(\textbf{k},\omega)$ is the distribution function column vector in Fourier space.

\begin{figure*}
\centering
\includegraphics[trim = 0 0 0 135,width=0.8\linewidth]{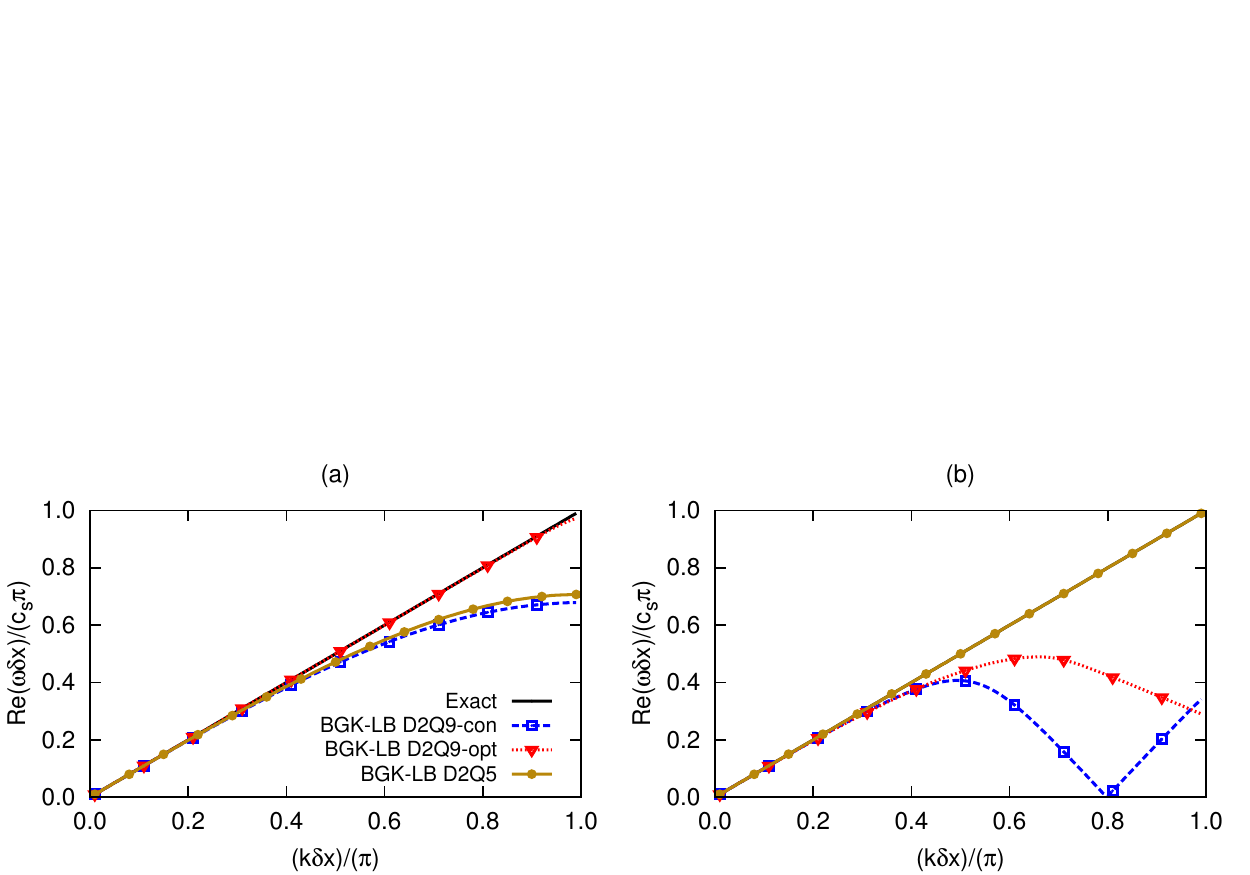} 
\caption{The optimum dispersion curves for the D2Q9 lattice. Here, the real component of the non-dimensional frequency $(\omega \delta x)/c_s\pi$, is plotted against the non-dimensional wave number $(k \delta x)/\pi$ for propagation along $\theta=0\degree$ shown in panel (a) and $\theta=45\degree$ shown in panel (b). Lattice weights used for D2Q5 are $w_0=0, w_1=1/4$. However, lattice weights required $w_0=0, w_1=0.001, w_5=0.249$ make the D2Q9 lattice equivalent to the D2Q5 lattice rotated by $45 \degree$. }
\label{fig:Optimised D2Q9 dispersion} 
\end{figure*}

The equation \cref{eq:MRT-LB dispersion} is solved numerically to obtain the linear MRT-LB dispersion relation. The eigenvector corresponding to the propagating mode is, in general, a linear combination of hydrodynamic variables pressure and velocity in moments space. In \cref{fig:MRT-LB dispersion}, dispersion curves for the BGK- and MRT-LB schemes on D2Q5 and D2Q9 lattices are plotted for propagation along $\theta=0\degree \textup{and}~\theta=45\degree$ (see \cref{fig:propagation direction}). The corresponding numerical attenuation (imaginary part of frequency $\omega$) is also plotted. The non-dimensional frequency (real and imaginary) $\omega^*=(\omega \delta x)/c_s\pi$ is plotted as a function of the non-dimensional wave-number $k^*=(k \delta x)/\pi$. Note that the non-dimensional wave-number $k^*$ also corresponds to grid resolution $R$ (Section \ref{sec:LB grid res}), with $k^*=1$ corresponding to the Nyquist limit of $2$ grid points per wavelength. Thus the working grid resolution for the LB scheme can be deduced from the dispersion curve by marking a point on the curve where it starts to deviate significantly from the exact curve. Lattice weights are determined from constraint equations \cref{eq:LB lattice moments,eq:0th lattice weight} with the condition that $\eta(\textbf{x})=1$ corresponding to a homogeneous medium. For the D2Q5 lattice, we obtain $w_0=0, w_1=w_2=w_3=w_4=1/4$. For the D2Q9 lattice, \cref{eq:LB lattice moments,eq:0th lattice weight} result in an under-determined set of equations for lattice weights. Thus, there is freedom to choose any combination of lattice weights for D2Q9 (but satisfying \cref{eq:LB lattice moments,eq:0th lattice weight}). For the curve in \cref{fig:MRT-LB dispersion} (D2Q9-con), we choose conventional lattice weights for D2Q9: $w_0=4/9, w_1=w_2=w_3=w_4=1/9, w_5=w_6=w_7=w_8=1/36$. 

Dispersion curves for the BGK- and MRT-LB scheme with $s_p = 2$ and $s_e = 2$ on the D2Q5 lattice are identical. Further, because the values of relaxation parameters for conserved moments, $s_\rho$ and $s_v$, are not relevant, the BGK- and MRT-LB  D2Q5 schemes we use are identical. Additionally, these schemes are attenuation free, as desired for our problem. The D2Q5 dispersion is exact for propagation along $\theta=45\degree$. The BGK-LB scheme on the D2Q9 lattice is also attenuation free. However, the dispersion curve deviates significantly from the exact, especially for propagation along $\theta=45\degree$. The MRT-LB scheme on the D2Q9 lattice gives better dispersion characteristics at the cost of (small) attenuation.  Also, these dispersion curves at best match D2Q5 counterparts. Thus the BGK-LB scheme on the D2Q5 lattice has the best dispersion characteristics.

Here we highlight the subtle distinction between free and forced waves which may be relevant for the dispersion analysis. Free waves have real wavenumber and complex frequency, whereas forced waves have complex wavenumber and real frequency. Hence free and forced waves respond differently to dissipation in the system and their dispersion characteristics are also affected \cite{kruger2017}. In our work, we are interested in comparing LB numerical schemes with standard finite-difference schemes. Hence, we are considering acoustic wave propagation in dissipation-free media for which the distinction between free and forced waves is not relevant \cite{lele1992compact}.

\subsection{Optimized lattice weights for the D2Q9 lattice}
For the D2Q5 lattice, equations \cref{eq:LB lattice moments,eq:0th lattice weight} yield a complete system of equations for lattice parameters. However, for the D2Q9 lattice, we obtain an under-determined system of equations. Since the D2Q9 lattice weights cannot be uniquely determined, we vary the lattice weights within these constraints. For each set of weights for the D2Q9 lattice, we calculate the numerical dispersion in equation \cref{eq:MRT-LB dispersion}. Performing this exercise, the lattice weights for the D2Q9 lattice which yield the best dispersion curves are obtained as $w_0=0, w_1=\varepsilon, w_5=1/4-\varepsilon$ where $\varepsilon$ is a very small positive number. Note that $\varepsilon=0$ causes instability, as the lattice sound speed $c_s$ becomes unity (see Appendix \ref{sec:Appendix D}). The optimum curves are plotted in \cref{fig:Optimised D2Q9 dispersion} (D2Q9-opt) and compared with the D2Q5 as well D2Q9 curves with conventional weights. We see that dispersion curves of D2Q9-opt match best with the exact curves for propagation along $\theta=0 \degree$. However, this particular choice of lattice weights makes D2Q9 roughly equivalent to D2Q5 rotated by $45 \degree$. Note that the D2Q9 lattice weights $w_0=0, w_1=1/4, w_5=0$ gives the D2Q5 lattice and hence identical dispersion relations to the D2Q5 scheme.
\subsection{Comparison with finite-difference schemes}
\begin{figure*}
\centering
\includegraphics[width=0.8\linewidth]{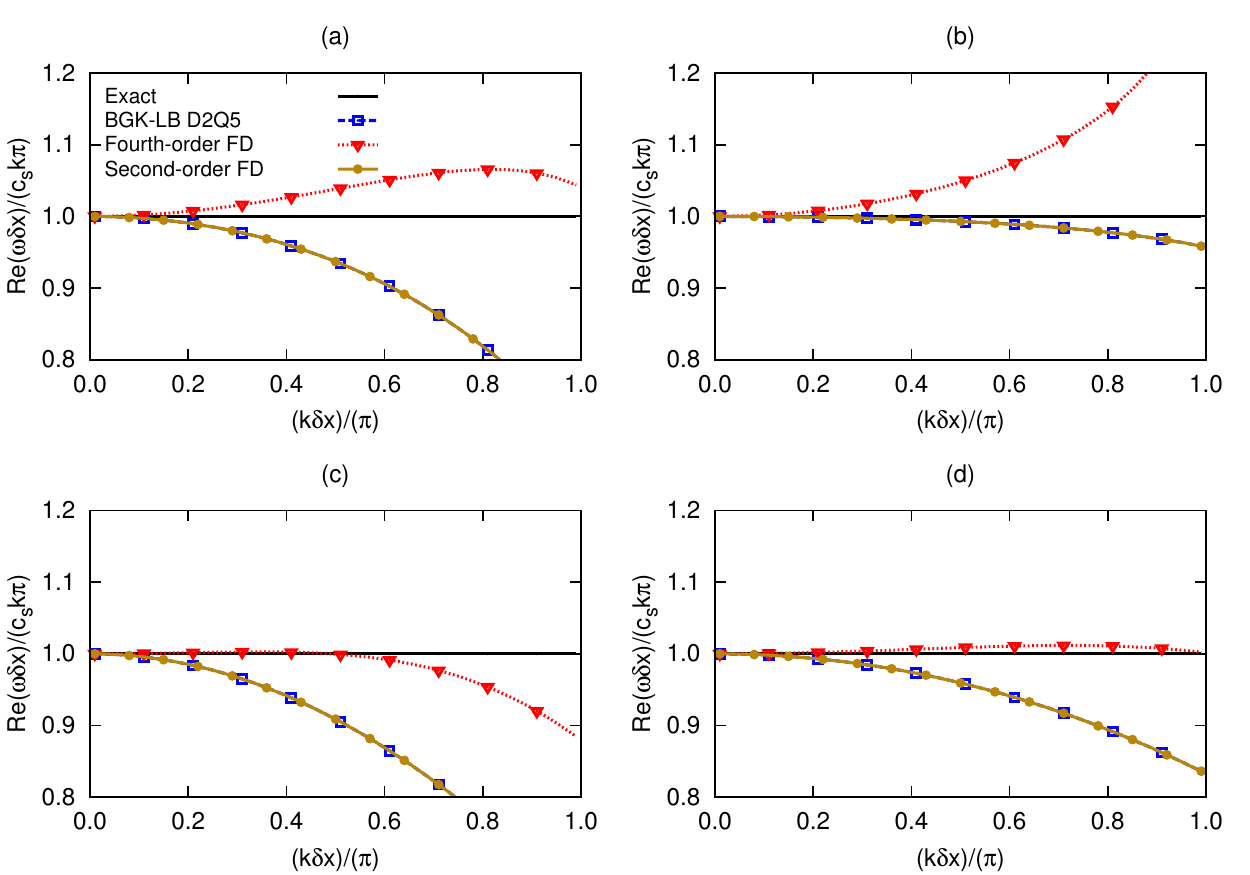} 
\caption{Comparison of the BGK-LB D2Q5 dispersion curves (normalised sound speed $(\omega \delta x)/k c_s\pi$ vs normalised wave-number $(k \delta x)/\pi$) with second- and fourth-order FD schemes. Panel (a) and (b) show numerical dispersion at Courant number $C=0.66$ (maximum allowed value for the fourth-order FD scheme) for propagation along $\theta=0\degree$ and $\theta=45\degree$  respectively. Panel (c) and (d) show  numerical dispersion at Courant number $C=0.33$ for propagation along $\theta=0\degree$ and $\theta=45\degree$ respectively. The BGK-LB D2Q5 curves are identical to the second-order FD curves for these propagation angles. The fourth-order FD scheme has better dispersion characteristics overall. The legend of the first plot also applies to the rest of the plots.}
\label{fig:Dispersion comparison 1} 
\end{figure*}
\begin{figure*}
\centering
\includegraphics[trim = 0 0 0 135, width=0.8\linewidth]{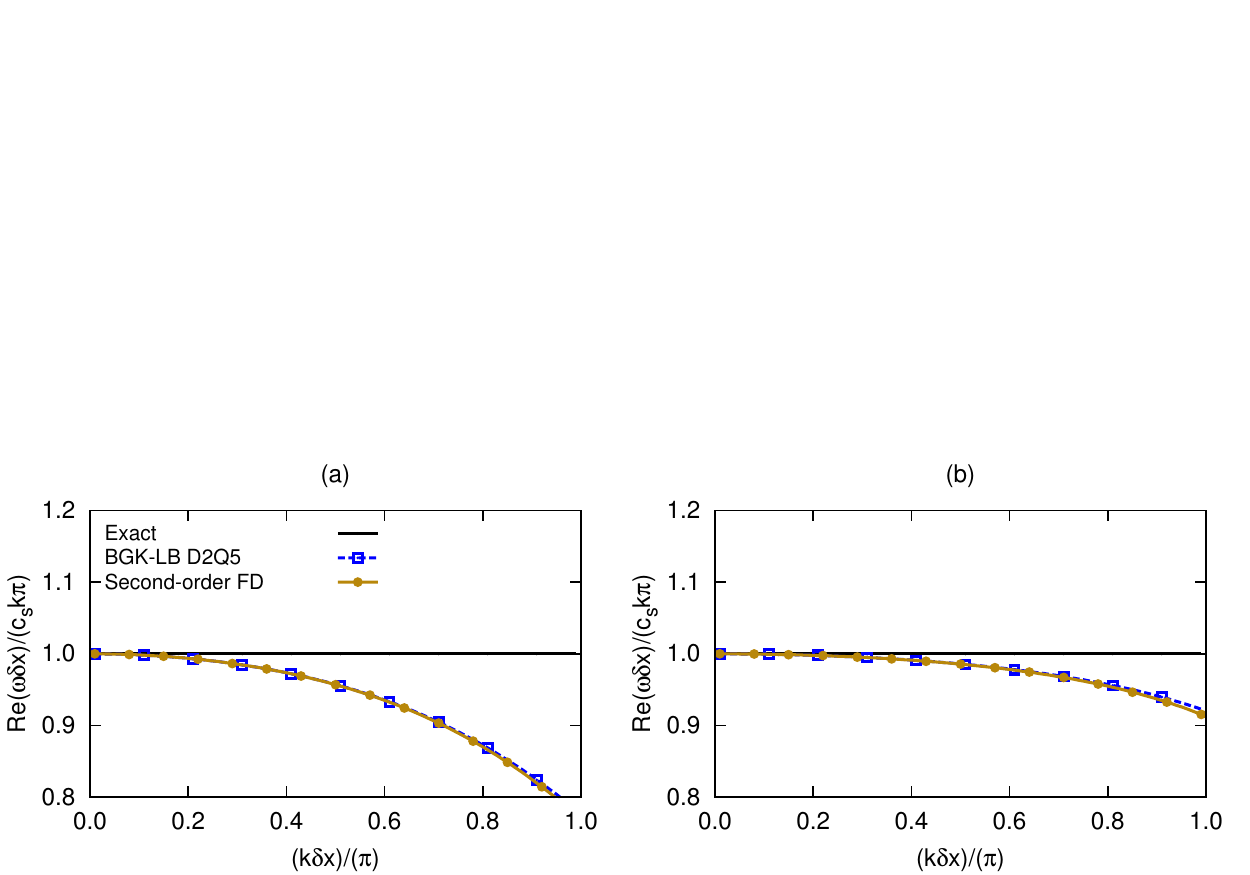} 
\caption{Comparison of the BGK-LB D2Q5 dispersion curves (normalised sound speed $(\omega \delta x)/k c_s\pi$ vs normalised wave-number $(k \delta x)/\pi$)  with exact and second-order FD scheme at Courant number $C=0.71$ (maximum allowed value for both the BGK-LB D2Q5 and the second-order FD scheme), for propagation along $\theta=15\degree$ shown in panel (a) and $\theta=30\degree$ shown in panel (b). These BGK-LB D2Q5 dispersion curves are slightly better than second-order FD. The legend of the first plot also applies to the second plot.}
\label{fig:Dispersion comparison 2} 
\end{figure*}
The discussion hitherto suggests that BGK-LB with D2Q5 is the best LB scheme for the simulation of linear acoustic waves. We compare this scheme with standard schemes used in seismology --- the second- and fourth-order finite-difference (FD) schemes. The stability of FD schemes as well as the LB scheme depends on the local Courant number which is given by \cite{robertsson2011numerical,Kupershtokh20102236} 
\begin{equation}\label{eq:LB Courant number 1}
C=c_s (\delta t/\delta x),
\end{equation}
where $c_s=c_s(\textbf{x})$ is the local sound speed. From equation \cref{eq:LB Courant number 1}, we see that the Courant number is ratio between the physical information propagation speed and the numerical information propagation speed. For the BGK-LB D2Q5 scheme, the numerical information can propagate only along the axes with speed $\delta x/\delta t = 1$ and hence it propagates with speed $1/\sqrt{2} \approx 0.71$ along the diagonal. Hence, the maximum Courant number $C_{max} \approx 0.71 $ which is also the maximum lattice sound speed (see Appendix \ref{sec:Appendix D}). For the second-order and the fourth-order FD scheme used here (see Appendix \ref{sec:Appendix C}), the maximum allowed Courant number $C_{max}$ is approximately equal to 0.71 and 0.66 respectively. Considering these limits, we study the dispersion relation for different values of the Courant number.
\begin{figure*}
  \subfloat[\label{fig:coarse and fine meshes}]{\includegraphics[width= 0.7\columnwidth]{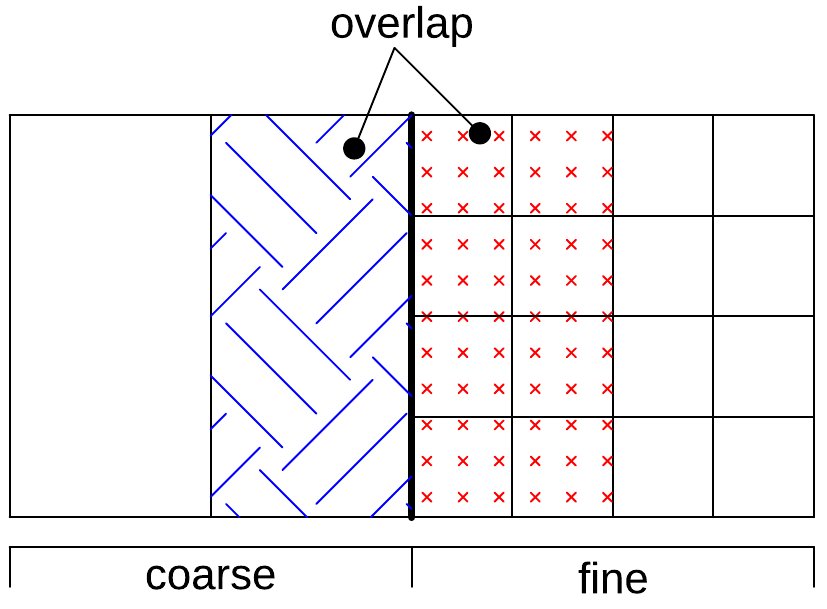}}
  \subfloat[\label{fig:algorithm flow}]{\includegraphics[width= 0.9\columnwidth]{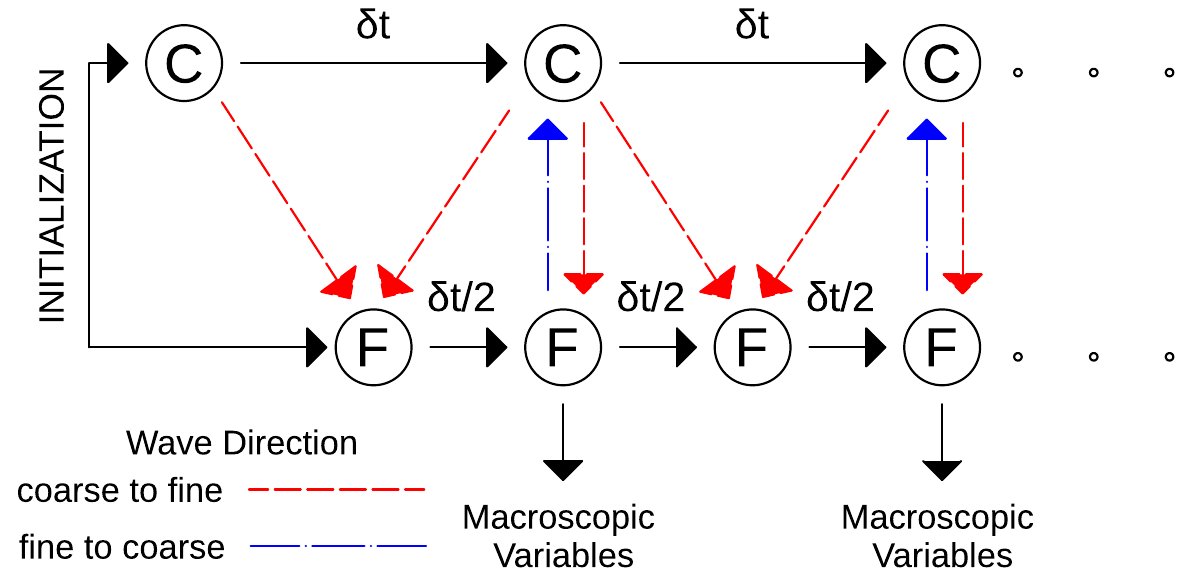}}
  \caption{\label{fig:grid refinement details}Mesh Refinement Scheme (a) Coarse-Fine Interface (b) LBM Sequence on composite mesh --- interpolation sequence is based on scheme proposed in \cite{PhysRevE.62.2219}}
\end{figure*}
In \cref{fig:Dispersion comparison 1}, the BGK-LB D2Q5 dispersion relation is compared with the second- and the fourth-order FD scheme at $C=0.66$ and $C=0.33$, for propagation along $\theta= 0\degree$ and $\theta= 45\degree$. These plots have normalised sound speed on the y-axis rather than normalised frequency. For these directions, the dispersion curves of the BGK-LB D2Q5 and the second-order FD scheme are identical (see Appendix \ref{sec:Appendix E}). As shown in \cref{fig:Dispersion comparison 2}, the BGK-LB D2Q5 and the second-order FD dispersion curves for the intermediate propagation angles are also identical, except at very low grid resolution where the BGK-LB D2Q5 is slightly better. The fourth-order FD scheme has better dispersion characteristics which further improve with decreasing Courant number. However, for propagation along $45\degree$ at high Courant numbers (close to $C_{max}=0.66$ for the fourth-order FD scheme), the BGK-LB D2Q5 (and the second-order FD) is more accurate.

\section{Grid Refinement} \label{sec:LB grid refinement}

Seismic waves travel through heterogeneous media where, background density and sound speed may vary significantly across the domain. The characteristic wavelength of seismic waves is directly proportional to the sound speed and the accuracy at low grid resolution (Section \ref{sec:LB grid res}) is an important performance criterion for numerical schemes modelling seismic waves. It is therefore desirable to keep the grid resolution, $R=(\lambda_c/\delta x)$, constant across the computational domain. If we model heterogeneous media as a patchwork of domains of homogeneous media, we will need a grid-refinement scheme for the LB simulation to maintain grid resolution across a  varying sound speed.

The underlying requirement for the LB grid-refinement algorithm is to facilitate streaming of distribution functions across different patches of locally homogeneous media. With the grid-refinement scheme, lattice spacing $\delta x$ is set to vary depending on the sound speed. To maintain the microscopic velocities, corresponding changes in the time step $\delta t$ are required \cite{PhysRevE.67.066707,PhysRevE.62.2219}. The LB algorithm naturally splits into two steps --- streaming and collision. So it is convenient to maintain a time-step ratio in multiples of $2$ between neighbouring patches of locally homogeneous media. Accordingly, the ratio of lattice spacing between two neighbouring patches of homogeneous media should also be in multiples of $2$.

Simulations in heterogeneous media were carried out using the LB grid-refinement scheme based on an algorithm suggested by Dupuis et al. \cite{PhysRevE.67.066707}. In order to understand the method, let us consider a heterogeneous medium consisting of two regions of homogeneous media connected at the interface (\cref{fig:coarse and fine meshes}). The region on the left has greater sound speed and thus larger characteristic wavelength than the region on the right. To maintain the grid resolution, we need a coarse grid on the left and a fine grid on the right. Let the ratio of lattice spacing between the two domains be $n=(\delta x_{c}/\delta x_{f})$ where $c$ stands for coarse and $f$ stands for fine. Note that $n$ is a multiple of $2$. During the interchange between the coarse and fine domains, distribution functions should be appropriately scaled. The distribution function can be composed as a sum of equilibrium and non-equilibrium parts. Since the equilibrium distribution is a function of local macroscopic variables $p$ and $\textbf{v}$, it should remain unaltered during the interchange. The scaling of the non-equilibrium part is determined through Chapman-Enskog analysis \cite{PhysRevE.67.066707}. Thus we have
\begin{equation}\label{eq:grid refinement eqm dist}
g_{i}^{eq,c}=g_{i}^{eq,f}=g_{i}^{eq},
\end{equation}
where $g_{i}^{eq,c}$ and $g_{i}^{eq,f}$ are the equilibrium distribution functions for the coarse and the fine domain respectively and
\begin{equation}\label{eq:coarse-fine dist connect}
\begin{split}
g_{i}^{c}=g_{i}^{eq}+(g_{i}^{f}-g_{i}^{eq}) n,\\
g_{i}^{f}=\hat{g}_{i}^{eq}+(\hat{g}_{i}^{c}-\hat{g}_{i}^{eq})(1/n),
\end{split}
\end{equation}
where $\hat{g}_{i}$ are spatially and temporally interpolated distribution functions on the coarse-grid. Since $\tau=1/2$ (BGK-LB scheme), the kinematic viscosity \cref{sec:LB viscosity} remains zero even if the time step $\delta t_l$ is varying across computational domain. As illustrated in \cref{fig:algorithm flow}, we follow the following procedure for time evolution on a composite coarse fine grid:
\begin{itemize}
\item Initialise coarse (C) and fine (F) grid distribution functions to equilibrium values at $t=0$.
\item Advance the coarse-grid distributions to $t=2$ with collision and streaming.
\item If the wave is travelling from the coarse to the fine domain, calculate the fine-grid distribution functions at the interface at $t=1$ (thick line) using \cref{eq:coarse-fine dist connect} with spatial and temporal interpolation of the coarse-grid distribution functions at $t=0$ and $t=2$. Else, fine-grid distribution functions are unchanged.
\item Evolve fine-grid distribution functions from $t=1$ to $t=2$ through collision and streaming.
\item If the wave is travelling from the coarse to the fine domain, correct fine-grid distribution functions at the interface according to \cref{eq:coarse-fine dist connect} using coarse-grid distribution functions only at $t=2$. Else, if the wave is travelling from the fine to the coarse domain, correct coarse-grid distribution functions at $t=2$ according to \cref{eq:coarse-fine dist connect} using fine-grid distribution functions at $t=2$.
\item Evaluate and register macroscopic variables at $t=2$ on all grid nodes.
\item Evolve the fine-grid distribution functions from $t=2$ to $t=3$.
\end{itemize}
The procedure is repeated for the desired total evolution time.

Attendant to simulating on non-uniform grids are spurious reflections that occur at the interface \cite{BAZANT1982451}. Indeed, the low order of accuracy of LB combined with the abrupt halving in grid spacing at the coarse-fine grid interface creates spurious reflections. Here we mitigate these spurious reflections by specifying the direction of wave propagation at the coarse-fine grid interface.

Another type of spurious reflection arises when distribution functions from the coarse region or the fine region hit the domain boundary at the interface. To counter it, we introduce an overlap region (see \cref{fig:coarse and fine meshes}) of few grid points where the coarse domain extends into the fine domain and vice versa. Streaming distributions coming from either domains, instead of stopping at the interface,  proceed in the overlap region on respective grids. The overlap regions consist nodes which form sponge layers \cite{Colonius2004}. At a sponge node, during each collision, the distribution function is reduced by a fraction of its value. Thus, the sponge layers gradually absorb the wave as it propagates in the overlap region. This results in negligible reflections back into the physical domain. However, physical reflections are not affected as accurate boundary conditions are still maintained at the interface. Macroscopic variables are evaluated using distribution functions at legitimate coarse and fine nodes.

\section{Numerical experiments} \label{sec:numerical experiments}
\subsection{Waves in homogeneous media}
\begin{figure}[b]
    \centering
    \includegraphics[width=0.75\columnwidth]{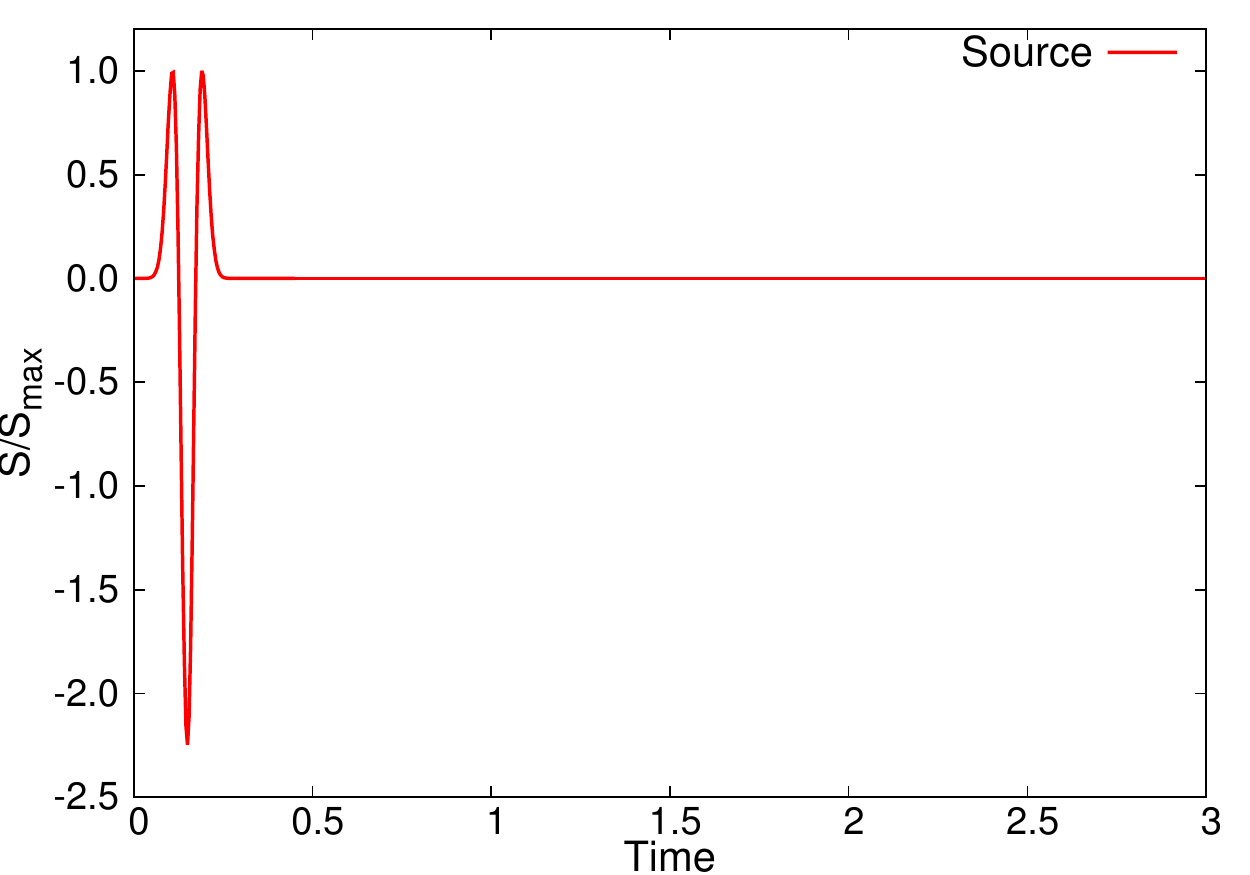} 
    \caption{Time variation of model source for acoustic simulations}
  \label{fig:source} 
\end{figure}
\begin{figure}
  \subfloat{\label{fig:simulation BGK-D2Q5} \includegraphics[width= 0.48\columnwidth]{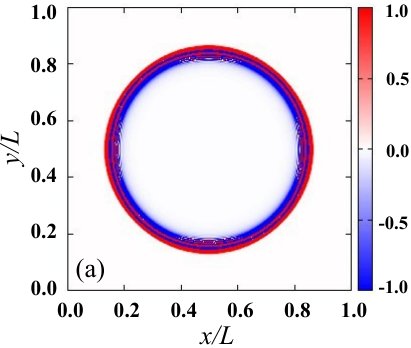}}
  \subfloat{\label{fig:simulation BGK-D2Q9} \includegraphics[width= 0.48\columnwidth]{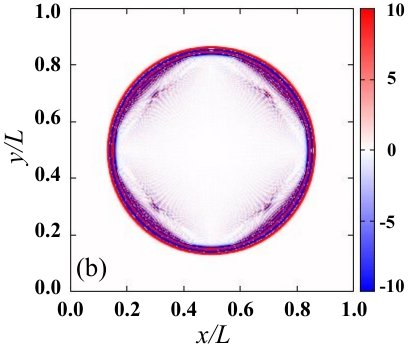}}\\
  \subfloat{\label{fig:simulation MRT-D2Q5} \includegraphics[width= 0.48\columnwidth]{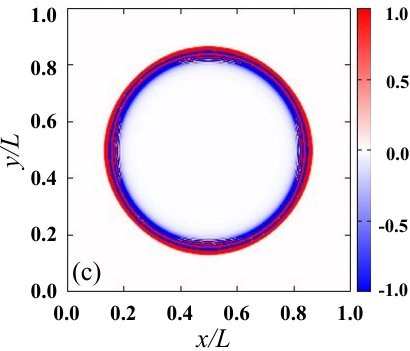}}
  \subfloat{\label{fig:simulation MRT-D2Q9} \includegraphics[width= 0.48\columnwidth]{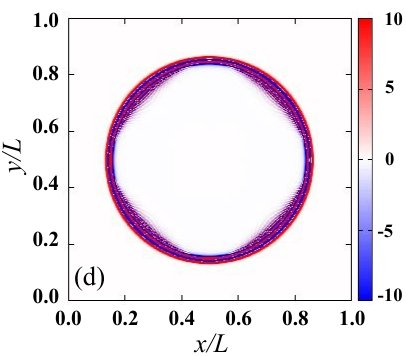}}
  \caption{\label{fig:LB wave simulation}Simulations of waves in homogeneous medium. Panel (a) and (b) show simulation using the BGK-LB scheme on D2Q5 and D2Q9 lattice respectively. Panel (c) and (d) show simulation using the MRT-LB scheme on D2Q5 and D2Q9 lattice respectively. The D2Q5 lattice with weights $w_{0}=0$, $w_{1}=1/4$ and the D2Q9 lattice with weights $w_{0}=0$, $w_{1}=0.01$ and $w_{5}=0.24$ are used. For the MRT-LB simulation $s_p,s_e=2$ on both lattices. For the D2Q9 lattice, $s_\epsilon$ is also set to $2$ and $s_q$ is set to $1$. The grid resolution is $16$ points per wavelength. The source is located at the center. Since the BGK-LB D2Q9 dispersion relation deviates significantly from exact, different wavenumbers propagate at different speeds, resulting in a series of trailing waves behind main wavefront. These trailing waves are eliminated in the MRT-LB simulation due to attenuation. Both the BGK-LB and the MRT-LB simulations on the D2Q5 lattice are identical and better than D2Q9.}
\end{figure}
\begin{figure*}
\centering
\includegraphics[width=1.6\columnwidth]{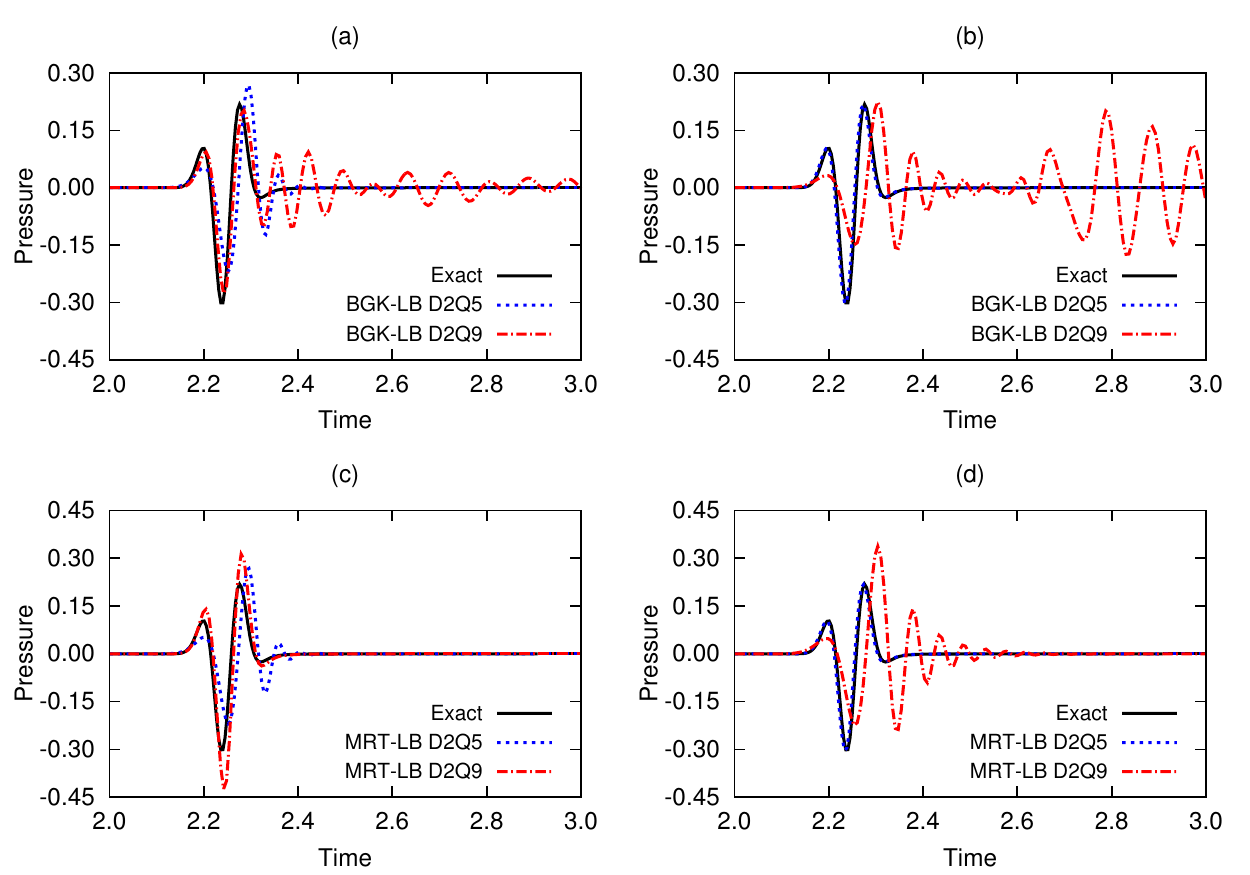} 
\caption{The time responses of the medium at distance $r=21\lambda$ from source. Panel (a) and (b) show the time response from BGK-LB simulation of wave propagation along $\theta=0 \degree$ and $\theta=45 \degree$ respectively. Panel (c) and (d) show the time response from MRT-LB simulation of wave propagation along $\theta=0 \degree$ and $\theta=45 \degree$ respectively. Numerical results are compared with the analytical result. Both BGK- and MRT-LB D2Q5 simulations are identical and the schemes capture the response pulse in close approximation to the analytical result. For the BGK-LB D2Q9 simulation, the primary pulse arrives approximately at the same time as the analytical result. However, there are a series of trailing response pulses after the main pulse. The MRT-LB D2Q9 simulation removes the trailing waves due to attenuation. But overall the performance is less accurate than D2Q5.} 
\label{fig:time response}
\end{figure*} 
The findings of the dispersion-relation analysis are reinforced through simulation tests in homogeneous media. For test simulations, the source used to mimic the seismic disturbance is
\begin{equation}\label{eq:source}
S(\textbf{x},t)=-(1-4\xi^{2})\exp(-2\xi^{2})\delta (\textbf{x}-\textbf{x}_s),
\end{equation}
where
\begin{equation}\label{eq:tau}
\xi=\frac{2\pi f_{c}}{3}\left(t-\frac{3}{2 f_{c}}\right),
\end{equation}
and where central frequency $f_c=10$ Hz and wave velocity $v_s(\textbf{x})=4$ km/s. The source is located at a point $\textbf{x}_s$ in the domain. The temporal variation of the source is shown in \cref{fig:source}.  For homogeneous media, we set the mean density $\rho_{0}(\textbf{x})=1$ kg/m$^{3}$. The domain size for the simulation is $60 \lambda_c$, $\lambda_c$ being the central wavelength and simulations are carried out with grid resolution varying from $16$ grid points per $\lambda_c$ to $4$ grid points per $\lambda_c$. For the source term given by \cref{eq:source}, wave equation \cref{eq:linearpresswave} admits an analytical solution (in Fourier space)
\begin{equation}\label{eq:exact solution}
\hat{p}(\textbf{x},\omega)=-\frac{\iota}{4}\frac{dS(\textbf{x},\omega)}{dt}H_0^{(2)}(k\lvert \textbf{x}-\textbf{x}_s \rvert),
\end{equation}
where $H_n^{(2)}$ is the Hankel function of the second kind. The solution in time domain is obtained by taking the inverse Fourier transform.
\begin{figure*}
    \centering
    \includegraphics[width=1.5\columnwidth]{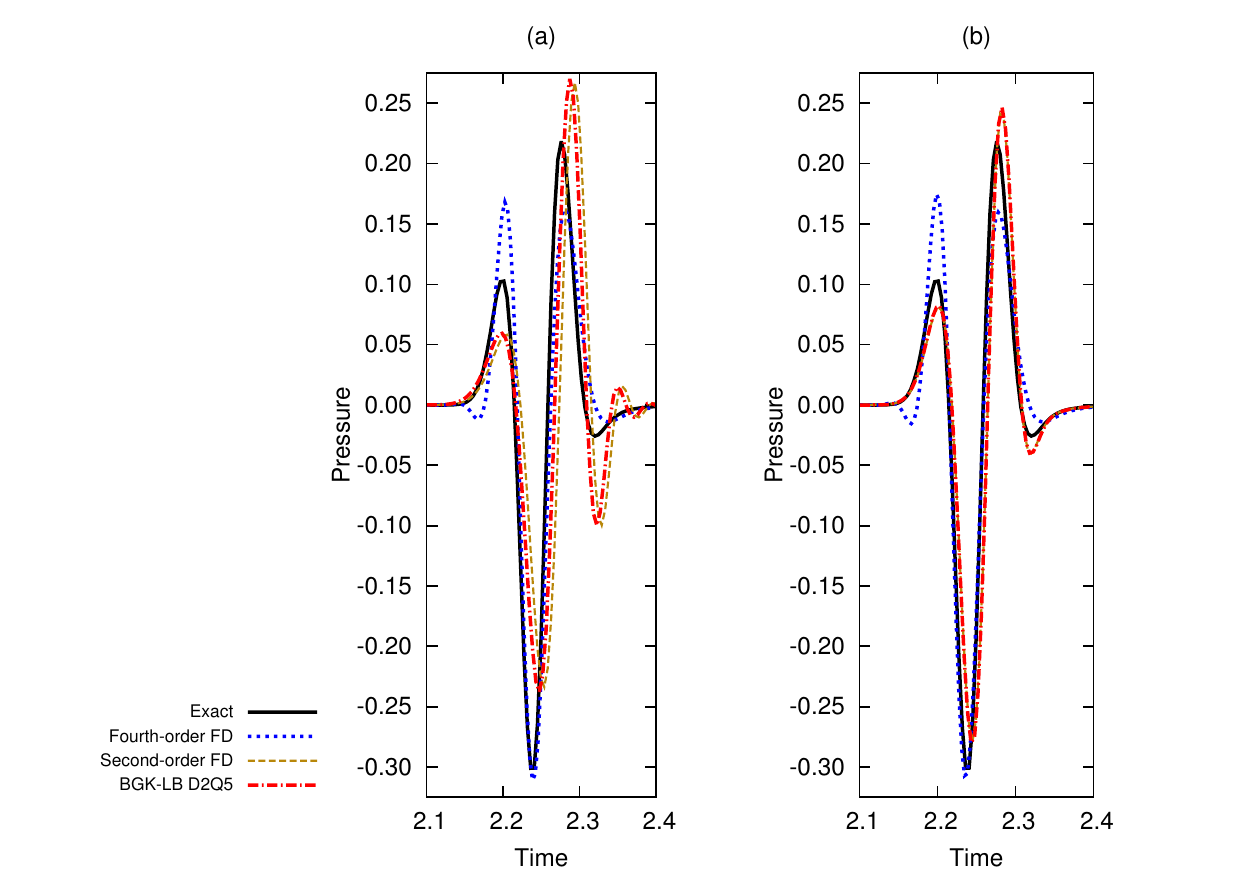} 
    \caption{ Comparison of the BGK-LB D2Q5 scheme with second- and fourth-order finite-difference schemes. Detailed view of time responses at distance $r=21\lambda$ from source for wave propagation along $\theta=15 \degree$  shown in panel (a) and $\theta=30 \degree$ shown in panel (b).}
  \label{fig:time response detailed} 
\end{figure*}
The wave is excited by the source \cref{eq:source} at a point $\textbf{x}_s$, typically at the center. The resulting waves propagate in all directions. \cref{fig:LB wave simulation} shows a snapshot of the propagating wavefront in the homogeneous medium. The LB simulations are carried out using the 2D lattices D2Q5 and D2Q9. In the BGK-LB D2Q5 and D2Q9 simulations, we use relaxation time $\tau = 1/2$ (see \cref{sec:LB parameters}). In the MRT-LB D2Q5 simulation we set only $s_p$ and $s_e$ equal to 2 and values of the other relaxation parameters are not relevant. In the MRT-LB D2Q9 simulation, $s_\epsilon$ is also set to 2 and $s_q$ is set to 1. For the D2Q5 lattice, simulations with the MRT-LB scheme  and the BGK-LB scheme are identical. For both  simulations, we see a distinct wavefront with a compact shape at $\theta=45 \degree $. For the BGK-LB D2Q9 simulation, we see a trail of waves following the main wavefront as the lattice responds differently to waves of different spatial wavenumbers. With the MRT-LB D2Q9 simulation, this trail of waves is eliminated as dispersion performance improves (see \cref{fig:MRT-LB dispersion}). However, the main wavefront in D2Q5 simulation is more isotropic and less dispersed than in D2Q9 simulation.

These results are compared to the exact analytical solution \cref{eq:exact solution} by studying the time response at different points in the medium. In \cref{fig:time response}, the time response curve at two points in the medium obtained using the BGK- and MRT-LB scheme on D2Q5 and D2Q9 lattices is compared with the exact result. The overall results for D2Q5 are better than for D2Q9, as expected from the dispersion analysis.

These results suggest that the BGK-LB D2Q5 is the best choice for simulating linear acoustic waves. A detailed look at the time response pulse and a comparison with second- and fourth-order finite-difference schemes for propagation along $\theta=15\degree$ and $\theta=30\degree$ is presented in \cref{fig:time response detailed}. We see that BGK-LB D2Q5 and the second-order finite-difference scheme results are comparable, consistent with the dispersion curves in \cref{fig:Dispersion comparison 2}. Overall however, the fourth-order finite-difference scheme is more accurate than the LB scheme. The simulations at low grid resolution show similar trends.

\subsection{Waves in heterogeneous media}
\begin{figure}
  \subfloat{\label{fig:Grid-refinement in homogeneous media} \includegraphics[width= 0.50\columnwidth]{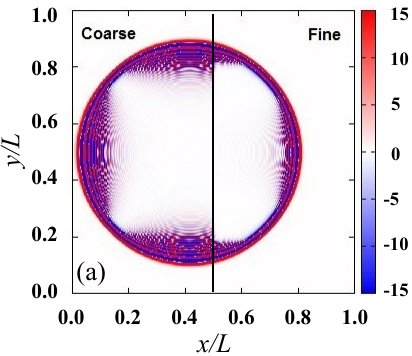}}
  \subfloat{\label{fig:Grid-refinement in heterogeneous media} \includegraphics[width= 0.50\columnwidth]{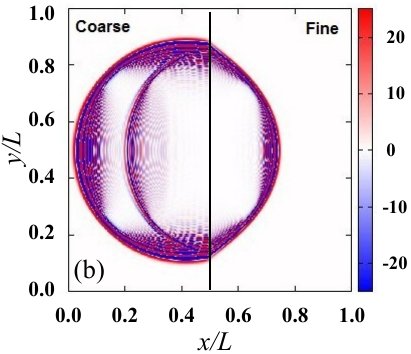}}
  \caption{\label{fig:Wave propagation in heterogeneous media}A snapshot of the propagating wavefront using grid refinement with the BGK-LB scheme on the D2Q5 lattice (a) in a homogeneous (uniform) medium (b) in a heterogeneous (non-uniform) medium. For both the media, the left side is the coarse computational domain and the right side is the fine computational domain. A line separating the two domains is also shown. For the homogeneous medium, we see no artificial reflection at the coarse-fine interface. For the heterogeneous medium, we see a reflection at the interface because of the sound speed contrast (0.8) between the coarse and fine domains.}
\end{figure}
The simulations in a heterogeneous medium are carried out using the same source \cref{eq:source}. The heterogeneous medium we consider comprises two homogeneous media with different sound speeds, joined together to form an interface. The ratio of sound speeds in the two domains, i.e velocity contrast, is 0.8. Accordingly, the grid on the left is coarse and the grid on the right is fine. We implement the grid-refinement algorithm discussed in Section \ref{sec:LB grid refinement}. The results are shown in \cref{fig:Wave propagation in heterogeneous media}.

\cref{fig:Grid-refinement in homogeneous media} shows simulations in a homogeneous medium with grid refinement. The source is located on the left i.e. coarse-grid. The BGK-LB scheme on the D2Q5 lattice is used. Waves propagate smoothly across the interface without any artificial reflection. The only difference between the left and the right is the resolution, which is $8$ points per wavelength on the coarse side and $12.8$ points per wavelength on the fine side.  Note that the simulated wavefront is broader in the coarse domain compared to \cref{fig:simulation BGK-D2Q5}. This is because resolution in the coarse domain is lowered by a factor of two compared to the resolution in the computational domain in \cref{fig:simulation BGK-D2Q5}. \cref{fig:Grid-refinement in heterogeneous media} shows the simulation in the heterogeneous medium. Waves from the coarse domain suffer a reflection when they cross the medium boundary. On the fine side, waves propagate with a reduced speed.

\section{Conclusions}
We have demonstrated a successful formulation of linear acoustic wave propagation using the BGK- and MRT-LB frameworks with a linear equilibrium distribution function, first proposed in \cite{Chopard1999115}. Similar to analysis performed in \cite{PhysRevE.61.6546} for the conventional LB scheme, we develop a formalism to calculate the dispersion relations for the linear BGK- and MRT-LB scheme. With our formalism, it is possible to compare the performance of various LB lattices for simulation of linear acoustic waves. Our formalism is also useful in comparing the dispersion relation of the LB schemes with standard finite-difference schemes for a given Courant number. The LB dispersion relations  as well as numerical simulations are in reasonable agreement with theoretical results.

Our dispersion analysis establishes that the fourth-order finite-difference scheme is better than any LB numerical scheme for the simulation of the linear acoustic waves. As with the finite-difference schemes \cite{lele1992compact}, the numerical dispersion of the LB scheme is also anisotropic. Our dispersion analysis shows that for the linear LB scheme, the numerical dispersion is most inaccurate if the wave is propagating along one of the directions of LB particle streaming.   Hence, the dispersion relations for the D2Q9 lattice with the BGK and the MRT schemes are worse than for the D2Q5 lattice. In general, the dispersion performance of the LB scheme is degraded as we increase the number of streaming directions on the lattice. For the D2Q5 lattice however, the BGK and the MRT dispersion performance is identical even with only two relaxation parameters $s_p$ and $s_e$ set equal to 2 in the MRT scheme, a requirement necessary to recover macroscopic equations \cref{eq:energy equation} and \cref{eq:momentum equation}. Thus, the BGK-LB scheme on the D2Q5 lattice is the best suited LB numerical model for seismic waves. Also, the BGK-LB D2Q5 scheme --- which is a second-order scheme --- is comparable with the second-order finite-difference scheme.

To keep the computational cost of simulation low, numerical scheme used should be sufficiently accurate even with low grid resolution i.e. the number of grid points per wavelength. For heterogeneous media, the wavelength in the domain changes as a function of the local sound speed. Depending on the change in wavelength, the grid size needs to be altered to maintain the number of grid points that resolve the wave i.e. the grid resolution. This requires a grid-refinement algorithm for the LB scheme to smoothly accommodate the change in the grid size across the computational domain. With the grid-refinement algorithm presented here, it is possible to maintain a uniform grid resolution and accuracy of the simulation over the entire computational domain. We have successfully implemented the algorithm to simulate waves in heterogeneous media  with velocity contrasts using the BGK-LB scheme on D2Q5 lattice. The techniques used here can be easily extended to 3D wave propagation problems in much more complicated environments.  Overall, the LB scheme is therefore a promising tool for faster, cost-effective simulation of waves in seismology. 

\section*{Acknowledgement}
We would like to thank TIFR for their support and the use of the SEISMO computer cluster which was used to perform the calculations presented here. SMH acknowledges support from Ramanujan fellowship SB/S2/RJN-73 2013, the Max-Planck Partner Group Program and the NYUAD Center for Space Science. We would also like to thank L-S Luo and J. O. Blanch for useful conversations. We thank the anonymous reviewers for their valuable comments and suggestions.

\appendix
\section{Transformation matrix $\textup{M}$} \label{sec:Appendix A}
For the D2Q5 lattice, 
\begin{align*}
\rm{M}=
  \begin{bmatrix}
    1 & 1 & 1 & 1 & 1 \\
    0 & 1 & 0 & -1 & 0 \\
    0 & 0 & 1 & 0 & -1 \\
    -4 & 1 & 1 & 1 & 1 \\
    0 & 1 & -1 & 1 & -1
  \end{bmatrix},
\end{align*}
and for the D2Q9 lattice, 
\begin{align*}
\rm{M}=
  \begin{bmatrix}
    1 & 1 & 1 & 1 & 1 & 1 & 1 & 1 & 1 \\
    0 & 1 & 0 & -1 & 0 & 1 & -1 & -1 & 1 \\
    0 & 0 & 1 & 0 & -1 & 1 & 1 & -1 & -1 \\
    -4 & -1 & -1 & -1 & -1 & 2 & 2 & 2 & 2 \\
    0 & 1 & -1 & 1 & -1 & 0 & 0 & 0 & 0 \\
    4 & -2 & -2 & -2 & -2 & 1 & 1 & 1 & 1 \\
    0 & -2 & 0 & 2 & 0 & 1 & -1 & -1 & 1 \\
    0 & 0 & -2 & 0 & 2 & 1 & 1 & -1 & -1 \\
    0 & 0 & 0 & 0 & 0 & 1 & -1 & 1 & -1
  \end{bmatrix}.
\end{align*}
\section{Chapman-Enskog analysis for the linear MRT-LB scheme on D2Q5 lattice} \label{sec:Appendix B}
In the Chapman-Enskog multiscale expansion procedure \cite{chapman1970,epic23739}, the distribution  function and the time and spatial derivatives are expanded in terms of the small expansion parameter $\epsilon$ (Knudsen number) \cite{ref1}.
\begin{equation} \label{eq:app1}
g_i=g_i^{(0)}+\epsilon g_i^{(1)}+ \epsilon^2 g_i^{(2)}+...,
\end{equation} 
\begin{equation} \label{eq:app2}
\partial_t=\epsilon \partial_t^{(1)}+ \epsilon^2 \partial_t^{(2)} \ \ \ \ \ \, \ \ \ \ \ \partial_x=\epsilon \partial_x^{(1)},
\end{equation} 
\begin{equation} \label{eq:app3}
\begin{aligned}[b]
g_{i}(\textbf{x}+\textbf{c}_{i}\delta t,&t+\delta	t)\\={}&\sum_{n=0}^{\infty}\frac{\epsilon^n}{n!}\big(\partial_t+\textbf{c}_i \cdot \nabla \big)^n g_{i}(\textbf{x},t).
\end{aligned}
\end{equation}
Using these expansions in \cref{eq:MRT-LB linear acoustic} (without the scalar source term) and retaining terms only upto $\mathcal{O}(\epsilon^2)$ we obtain following equations (order by order in $\epsilon$):
\begin{equation} \label{eq:apporder0}
\epsilon^0: g_i^{(0)}=g_i^{eq},
\end{equation}

\begin{widetext}
\begin{equation} \label{eq:apporder1}
\epsilon^1: (\partial_t^{(1)}+\textbf{c}_i \cdot \nabla^{(1)}) g_i^{(0)}=-(1/\delta t) \sum_{j}\textit{C}_{ij}g_{j}^{(1)}=-(1/\delta t)\sum_{j}(\textup{M}^{-1}~\textup{S}~\textup{M})_{ij}g_{j}^{(1)},
\end{equation}
\begin{equation} \label{eq:apporder2}
\begin{aligned}[b]
\epsilon^2: \partial_t^{(2)} g_i^{(0)}+(\partial_t^{(1)}+\textbf{c}_i \cdot \nabla^{(1)}) g_i^{(1)}&+(\delta t/2)(\partial_t^{(1)}+\textbf{c}_i \cdot \nabla^{(1)})^2 g_i^{(0)} \\={}& -(1/\delta t)\sum_{j}\textit{C}_{ij}g_{j}^{(2)}=-(1/\delta t)\sum_{j}(\textup{M}^{-1}~\textup{S}~\textup{M})_{ij}g_{j}^{(2)}.
\end{aligned}
\end{equation}
\end{widetext}

Transforming these equations into the moment space gives
\begin{equation} \label{eq:apporder0moments}
\epsilon^0: \textbf{m}^{(0)}= \textbf{m}^{eq},
\end{equation}
\begin{equation} \label{eq:apporder1moments}
\begin{aligned}[b]
\epsilon^1: (\partial_t^{(1)} \textup{I}+\textbf{\textit{E}} &\cdot \nabla^{(1)})~\textbf{m}^{(0)}\\={}&-(1/\delta t) \textup{S}~\textbf{m}^{(1)},
\end{aligned}
\end{equation}
\begin{equation} \label{eq:apporder2momentsprime}
\begin{aligned}[b]
\epsilon^2: &\ \partial_t^{(2)} \textbf{m}^{(0)}+(\partial_t^{(1)} \textup{I}+\textbf{\textit{E}} \cdot \nabla^{(1)})  \textbf{m}^{(1)}\\&+(\delta t/2)(\partial_t^{(1)} \textup{I}+\textbf{\textit{E}} \cdot \nabla^{(1)})^2 ~ \textbf{m}^{(0)}\\&={}-(1/\delta t) \textup{S}~\textbf{m}^{(2)},
\end{aligned}
\end{equation}
where $\textbf{\textit{E}}=\textup{M}~\textup{diag}(\textbf{c}_0,\textbf{c}_1,\textbf{c}_2,...)~\textup{M}^{-1}$($\textup{diag}$ refers to diagonal matrix) and $\textup{I}$ is the Identity matrix. Equation \cref{eq:apporder2momentsprime} can be simplified using \cref{eq:apporder1moments} as
\begin{equation} \label{eq:apporder2moments}
\begin{aligned}
\epsilon^2: \partial_t^{(2)} \textbf{m}^{(0)}&+(\partial_t^{(1)} \textup{I}+\textbf{\textit{E}} \cdot \nabla^{(1)})(\textup{I}-\textup{S}/2) \textbf{m}^{(1)}\\&={}-\textup{S}~\textbf{m}^{(2)}.
\end{aligned}
\end{equation}

In the case of the D2Q5 lattice we have
\begin{align*}
\textup{M}^{-1}=
  \begin{bmatrix}
    0.2 & 0.0 & 0.0 & -0.2 & 0.0 \\
    0.2 & 0.5 & 0.0 & 0.05 & 0.25 \\
    0.2 & 0.0 & 0.5 & 0.05 & -0.25 \\
    0.2 & -0.5 & 0.0 & 0.05 & 0.25 \\
    0.2 & 0.0 & -0.5 & 0.05 & -0.25
  \end{bmatrix},
\end{align*}
also
\begin{align*}
\textup{E}_{\textup{x}}=
  \begin{bmatrix}
    0 & 1 & 0 & 0 & 0 \\
    0.4 & 0 & 0 & 0.1 & 0.5 \\
    0 & 0 & 0 & 0 & 0 \\
    0 & 1 & 0 & 0 & 0 \\
    0 & 1 & 0 & 0 & 0
  \end{bmatrix},
\end{align*}
\begin{align*}
\textup{E}_{\textup{y}}=
  \begin{bmatrix}
    0 & 0 & 1 & 0 & 0 \\
     0 & 0 & 0 & 0 & 0 \\
    0.4 & 0 & 0 & 0.1 & -0.5 \\   
    0 & 0 & 1 & 0 & 0 \\
    0 & 0 & -1 & 0 & 0
  \end{bmatrix}.
\end{align*}
Using \cref{eq:linear eqm dist}, components of the equilibrium moment vector are obtained as
\begin{align*}
\textbf{m}^{eq}=
  \begin{bmatrix}
    \rho^{(0)} \\
    \rho_0 v_x^{(0)} \\
    \rho_0 v_y^{(0)} \\   
    e^{(0)} \\
    p_{xx}^{(0)} 
  \end{bmatrix} =
  \begin{bmatrix}
    \rho(\textbf{x},t) \\
    \rho_0(\textbf{x}) v_x(\textbf{x},t) \\
    \rho_0(\textbf{x}) v_y(\textbf{x},t) \\   
    -4\rho + 10\rho(\textbf{x},t) c_s^2(\textbf{x}) \\
    0
  \end{bmatrix}.
\end{align*}
For the conserved moments --- fluctuating density and momentum --- $\rho^{(k)},\rho_0 v_x^{(k)},\rho_0 v_x^{(k)}=0$ for $k>0$. Writing $\rho=\rho(\textbf{x},t)$, $\textbf{v}=\textbf{v}(\textbf{x},t)$, $\rho_0= \rho_0(\textbf{x})$ and $c_s=c_s(\textbf{x})$ and substituting these values in \cref{eq:apporder1moments}, we obtain at the $\mathcal{O}(\epsilon)$ following set of equations for each component of $\textbf{m}$,
\begin{equation} \label{eq:appceo1c0}
m_0:\partial_t^{(1)} \rho+\nabla^{(1)} \cdot (\rho_0\textbf{v})=0,
\end{equation}
\begin{equation} \label{eq:appceo1c12}
m_1 + m_2:\partial_t^{(1)} (\rho_0\textbf{v})+\nabla^{(1)}(\rho c_s^2)=0,
\end{equation}
\begin{equation} \label{eq:appceo1c3}
\begin{aligned}[b]
m_3:\partial_t^{(1)} (10\rho c_s^2) &+ 5\ \nabla^{(1)} \cdot (\rho_0\textbf{v})\\&={}-(s_e/\delta t) e^{(1)},
\end{aligned}
\end{equation}
\begin{equation} \label{eq:appceo1c4}
m_4:\partial_x^{(1)}(\rho v_x)-\partial_y^{(1)}(\rho_0 v_y)=-(s_p/\delta t) p_{xx}^{(1)}.
\end{equation}
Similarly using \cref{eq:apporder2moments}, we obtain at $\mathcal{O}(\epsilon^2)$, for the conserved components of $\textbf{m}$ (using $e^{(1)}$ and $p_{xx}^{(1)}$ from \cref{eq:appceo1c3,eq:appceo1c4} respectively)
\begin{equation} \label{eq:appceo2c0}
m_0:\partial_t^{(2)} \rho=0,
\end{equation}

\begin{widetext}
\begin{align} \label{eq:appceo2c12}
\begin{aligned}[b]
m_1 + m_2:&\ \partial_t^{(2)} (\rho_0\textbf{v})-\delta t (c_s^2 -1/2) (1/s_e-1/2) \nabla^{(1)} \left[\partial_t^{(1)} (10\rho c_s^2) + 5\ \nabla^{(1)} \cdot (\rho_0\textbf{v})\right] \\&- (\delta t/2)(1/s_p-1/2) \left\{\partial_x^{(1)} \left[\partial_x^{(1)}(\rho v_x)-\partial_y^{(1)}(\rho_0 v_y)\right] \hat{\textbf{x}}+\partial_y^{(1)} \left[\partial_x^{(1)}(\rho v_x)-\partial_y^{(1)}(\rho_0 v_y)\right] \hat{\textbf{y}} \right\}=0.
\end{aligned}
\end{align}
\end{widetext}
\cref{eq:appceo1c0,eq:appceo2c0} add to give the continuity equation \cref{eq:energy equation} (without the scalar source term). Since the D2Q5 lattice cannot satisfy isotropy of the fourth order tensor \cite{frisch1986lattice,epic23739}, the exact recovery of the Navier-Stokes equation is not possible from \cref{eq:appceo1c12,eq:appceo2c12}. However, by setting $s_p=2$ and either $s_e=2$ or $c_s^2=1/2$, the linear conservation of momentum \cref{eq:momentum equation}, is recovered.

\section{Finite-difference schemes used for comparison} \label{sec:Appendix C}
The second-order accurate (O[2,2]) finite-difference solution of the linear wave equation is obtained by using centered differences for the spatial and temporal dervatives, 
\begin{equation}\label{eq:second-order leapfrog}
\begin{aligned}[b]
(u_i^{n+1}-2u_i^{n}&+u_i^{n-1})/\Delta t^2\\&={}c^2(u_{i+1}^{n}-2u_i^{n}+u_{i-1}^{n})/\Delta x^2.
\end{aligned}
\end{equation}
The fourth-order accurate (O[2,4]) solution is obtained by solving coupled Eqs. \cref{eq:energy equation,eq:momentum equation} simultaneously. These equations are of the general form
\begin{equation}\label{eq:simultaneous first-order}
\partial_t u=a\partial_x v, \ \ \ \ \ \ \ \ \partial_t v=b \partial_x u.
\end{equation}
We use a 4-point stencil for spatial derivatives as specified below.
\begin{equation}
\begin{aligned}[b]
&(u_i^{n+1}-u_i^{n})/\Delta t\\&={}a(-v_{i+1}^{n}+27v_{i}^{n}-27v_{i-1}^{n}+v_{i-2}^{n})/24\Delta x,
\end{aligned}
\end{equation}
\begin{equation}
\begin{aligned}[b]
&(v_i^{n+1}-v_i^{n})/\Delta t\\&={}b(-u_{i+2}^{n+1}+27u_{i+1}^{n+1}-27u_{i}^{n+1}+u_{i-1}^{n+1})/24\Delta x.
\end{aligned}
\end{equation}
Note that we can also derive equation \cref{eq:second-order leapfrog} from equation \cref{eq:simultaneous first-order} by taking second-order centered differences and eliminating $v$. We stop the simulation well before the waves reach boundary of the computational domain.

\section{The Courant number for the LB scheme} \label{sec:Appendix D}
The local Courant number given by \cref{eq:LB Courant number 1} depends on the local sound speed $c_s$. For the LB scheme, the dependence of $C$ on lattice parameters can be deduced from \cref{eq:LB lattice moments}. In particular, for the D2Q5 lattice, the relation between $C$ and the rest particle weight $w_0$ is given by
\begin{equation}\label{eq:app LB Courant number 2}
C=\sqrt{\frac{1-w_0}{2}}.
\end{equation}
The rest particle weight $w_0$ is modulated in order to obtain the locally varying lattice sound speed for heterogeneous media. However, it is essential to choose $w_0$ such that $C$ is always real and less than unity which is always true for the D2Q5 lattice. For homogeneous media, it is convenient to set $w_0=0$ which gives maximum allowed Courant number $C_{max}=1/\sqrt{2}$ for the LB D2Q5 scheme. In case of D2Q9, there is no definite relation between $C$ and $w_0$. Hence, $w_0$ needs to be chosen carefully.

\section{The analytical 1-D dispersion relation for the MRT-LB D2Q5 scheme} \label{sec:Appendix E}
\begin{figure}[ht]
\centering
\includegraphics[scale=0.325]{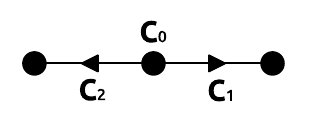}
\caption{The D1Q3 lattice: Projection of the D2Q5 lattice in 1-D}
\label{fig:latticesA}
\end{figure}

The solution of the eigenvalue equation \cref{eq:MRT-LB dispersion} gives the dispersion relation for the MRT-LB scheme. For the D2Q5 lattice, the characteristic equation is a fifth-order polynomial and hence writing down the general analytical solution is not possible. However, one can obtain the analytical solution for the propagation along the x-axis ($\theta = 0\degree$) by considering projection of the D2Q5 lattice on the x-axis i.e. the D1Q3 lattice (see \cref{fig:latticesA}).

In the D1Q3 lattice, the lattice velocities are $c_0 = (0,0)$ and $c_{1,2} = (\pm 1,0)$. Using the constraints on the lattice weights (equation \cref{eq:LB lattice moments}) and the lattice symmetry we have, $w_1=w_2=w$ and $w_0=1-2w$. Also the lattice sound speed $c_s=2w$ (homogeneous media). The matrix
\[
\rm{M}=
  \begin{bmatrix}
    1 & 1 & 1 \\
    0 & 1 & -1 \\
   -2 & 1 & 1 
  \end{bmatrix}
\]
transforms distribution functions to the moments space variables $\textbf{m}=\{\rho,\rho_0 v_x, e\}$. For the D1Q3 lattice, the matrix $\textrm{A}$ in equation \cref{eq:MRT-LB vector} is
\[
\rm{A}=
  \begin{bmatrix}
    2w & 2w-1 & 2w-1 \\
    -w & 1-w(1+1/c_s^2) & -w(1-1/c_s^2) \\
    -w & -w(1-1/c_s^2) & 1-w(1+1/c_s^2)
   \end{bmatrix}.
\] Substituting this in the eigenvalue equation \cref{eq:MRT-LB dispersion} yields the characteristic equation
\begin{equation} \label{eq:appD1Q3_characteristic polynomial}
 \left(\lambda + 1 \right) \left[\lambda^2-\lambda\  2\left(1-2c_s^2\sin^2(k/2) \right)+1 \right]=0,
\end{equation} where $\lambda = \exp(-i \omega)$. Clearly, one root is $\lambda = -1$ and hence $\omega = \pi$ which corresponds to a non-propagating mode. The other roots are
\begin{equation} \label{eq:appD1Q3_other roots}
\begin{aligned}[b]
\lambda ={}& \left(1-2c_s^2\sin^2(k/2) \right) \\& \pm i \sqrt{1-\left(1-2c_s^2\sin^2(k/2) \right)^2}.
\end{aligned}
\end{equation}
With the choice $\left(1-2c_s^2\sin^2(k/2) \right) = \cos{\theta}$, $\lambda = \exp{(\pm i \theta)}$ and eigenvalues for the propagating modes
\begin{equation} \label{eq:appD1Q3_dispersion}
\begin{aligned}[b]
\omega ={}& \pm \arccos{\left(1-2c_s^2\sin^2(k/2) \right)}\\={}& \pm 2 \arcsin{\left(c_s.\sin{(k/2)}\right)}.
\end{aligned}
\end{equation} This is identical to the 1-D dispersion relation for the second-order Fd scheme in \cref{eq:second-order leapfrog}.

%
\end{document}